\documentclass[pre,twocolumn,superscriptaddress]{revtex4}
\usepackage{graphicx}
\usepackage{float}
\usepackage{flafter}
\usepackage{float}
\usepackage{epsfig}
\usepackage{amsmath, amsthm, amssymb}

\def\gapx{\ \lower 2pt \hbox{$\buildrel>\over{\scriptstyle{\sim}}$}}
\def\lapx{\ \lower 2pt \hbox{$\buildrel<\over{\scriptstyle{\sim}}$}}
\def\lap{\ \lower 2pt \hbox{$\buildrel<\over{\scriptstyle{\sim}}$}}
\def\2d{$2\Delta$}
\def\3he{$^3$He}
\def\4he{$^4$He}
\def\A{\AA$^{-1}$}

\def\a1{$a_{1}$}
\def\a2{$\overline\alpha_{2}$}

\def\Am3{\AA$^{-3}$}
\def\ax2{$a_{2}$}

\def\dsq{$\Delta ^2 (t)$}
\def\d2o{D$_2$O}

\def\h2o{H$_2$O}

\def\Ke3{$\langle{K_3}\rangle$}
\def\k2{$\langle{[k_{\alpha}]^2}\rangle$}

\def\sqwe{$S(\textbf{Q},\omega~=0)$}
\def\srqwe{$S_R(\textbf{Q},\omega~=0)$}
\def\srqw{$S_R(\textbf{Q},\omega)$}
\def\sqw{$S(\textbf{Q},\omega)$}

\def\s1{$S_1(Q,E)$}

\def\sqw{$S(\textbf{Q},\omega)$}

\def\tr{$\tau_R$}

\def\m20{$M=20$}
\def\d2o{$D_{2}O$}

\def\a2{$\AA^2$}

\def\cn2{$\alpha$}

\def\Iiqt{$I_{inc}(\textbf{Q},t)$}
\def\iqt{$I(\textbf{Q},t)$}
\def\ii{$I_{\infty}$}

\def\om{$\omega$}

\def\rs{$\langle r^2 \rangle$}
\def\rsr{$\langle r^2\rangle_{R}$}
\def\rsmd{$\langle r^2\rangle_{MD}$}

\def\rsslope{$\langle r^2 \rangle_{R}$}

\def\rssim{$\langle \Delta ^2(t) \rangle /2$}

\def\td{$T_D$}

\def\tse{$T\simeq 100$}

\def\t150{$T\sim150$}

\def\tse2{$T\simeq 250$}

\def\u0{$(u=0)$}

\def\w2l{$\omega^2=\omega _L^2=\phi_L/M$}

\def\x0{$x=0$}

\begin{document}

\title{Long-Time Mean Square Displacements in Proteins}
\author{Derya Vural}                                                                                          
\affiliation{Department of Physics and Astronomy, University of Delaware, Newark, Delaware 19716-2570, 
USA}
\author{Liang Hong}                                                                                          
\affiliation{Center for Molecular Biophysics, Oak Ridge National Laboratory, Oak Ridge, TN 37831, USA}
\affiliation{Department of Biochemistry and Cellular and Molecular Biology, University of Tennessee, Knoxville, TN 37996, USA}
\author{Jeremy C. Smith}                                                                                          
\affiliation{Center for Molecular Biophysics, Oak Ridge National Laboratory, Oak Ridge, TN 37831, USA}
\affiliation{Department of Biochemistry and Cellular and Molecular Biology, University of Tennessee, Knoxville, TN 37996, USA}
\author{Henry R. Glyde}                                                                                         
\affiliation{Department of Physics and Astronomy, University of Delaware, Newark, Delaware 19716-2570, USA}     

\date{\today}

\begin{abstract}

We propose a method for obtaining the intrinsic, long time mean square displacement (MSD) of atoms and molecules in proteins from finite time molecular dynamics (MD) simulations. Typical data from simulations are limited to times of 1 to 10 ns and over this time period the calculated MSD continues to increase without a clear limiting value. The proposed method consists of fitting a model to MD simulation-derived values of the incoherent intermediate neutron scattering function, \Iiqt, for finite times. The infinite time MSD, \rs, appears as a parameter in the model and is determined by fits of the model to the finite time \Iiqt. Specifically, the \rs~ is defined in the usual way in terms of the Debye-Waller factor as $I(\textbf{Q},t = \infty) = \exp(-   Q^2 \langle\ r^2 \rangle /3)$. The method is illustrated by obtaining the intrinsic MSD \rs~ of hydrated lysozyme powder (h = $0.4 $ $g$ water/$g$ protein) over a wide temperature range. The intrinsic \rs~ obtained from data out to 1 ns and to 10 ns is found to be the same. The intrinsic \rs~ is approximately twice the value of the MSD that is reached in simulations after times of 1 ns which correspond to those observed using neutron instruments that have an energy resolution width of 1 $\mu$eV.  

\end{abstract}
\maketitle

\section{Introduction}

The mean-square displacement (MSD) of an atom is a fundamental dynamical quantity. In proteins the temperature dependence of the average MSD has been widely used to characterize the internal flexibility of the protein.\cite{Doster:89,Zaccai:00,Fitter:96,Chen:06,Roh:05,Roh:06,Gabel:02,Tarek:00,Chen:10,Daniel:98,Dunn:00} The MSD can be extracted from dynamic neutron scattering experiments. Since the incoherent neutron scattering cross section of the hydrogen (H) nucleus is large, the observed MSD is dominated by the MSD of H in the protein. As a result, the MSD of hydrogen in proteins has been extensively investigated by neutron scattering\cite{Doster:89,Zaccai:00,Fitter:96,Daniel:03,Chen:06,Roh:05} and the results compared with molecular dynamics simulation, a technique particularly complementary to dynamic neutron scattering.\cite{Smith:91} These studies have been performed as a function of temperature, pressure, hydration and in a variety of solvents.\cite{Rupley:91}

Most neutron and MD  studies to date have investigated the MSD on a picosecond-nanosecond (ps-ns) timescale.\cite{Doster:89,Doster:90,Ferrand:93,Jasnin:10,Nakagawa:10,Roh:05,Smith:90,Hayward:02,Hayward:03,Hayward:03a,Hamon:06,Calandrini:08,Calandrini:08a,Dirama:05,Dirama:06,Lerbret:08} At low temperatures ($T < 100$ K), a protein is essentially harmonic. As the temperature is increased beyond the harmonic regime, proline puckering transitions and methyl rotations are activated.\cite{Miao:12} These onsets are independent of protein hydration. At temperatures $T \simeq$ 160 - 220 K, jumps of non-exchangeable hydrogen in the non-proline methylene groups and aromatic phenyl rings dominate the MSD on the $10^{-9}$ second time scale. For $T \le$ 220 K protein flexibility arises chiefly from the hydrophobic and aromatic residues. In contrast, motion of the hydrophilic residues remains suppressed due to stable hydrogen bonding interactions with the neighboring protein residues and hydration water. As $T$ is further increased, at $T_D \sim 180-220$ K, a strongly hydration-dependent increase is found in the localized diffusion of protein non-exchangeable hydrogen atoms and in jumps in the hydrophilic side chains. The resulting increase in MSD is denoted the dynamical transition (DT). The jumps in hydrophilic side chains are strongly coupled to the relaxation rates of the hydrogen bonds formed with hydration water.\cite{Miao:12}

MD has also been used to probe the pressure dependence of protein MSDs.\cite{Calandrini:08,Calandrini:08a,Meinhold:05,Meinhold:07} These MD studies revealed a qualitative change in the internal protein motions at $p \sim 4$ kbar and the existence of two linear regimes in the MSD. The qualitative change is a loss with increasing pressure of large amplitude, collective protein modes below 2 THz effective frequency, accompanied by restriction of large-scale solvent translational motion.\cite{Meinhold:05} The DT was found to be pressure-independent, indicating that the effective energy barriers separating conformational substates are not significantly influenced by pressure. In contrast, vibrations within substates stiffen with pressure, due to increased curvature of the local harmonic potential in which the atoms vibrate.\cite{Meinhold:07}

Given the extensive interest in MSDs, both observed and simulated, it is useful to clarify in detail what is measured and calculated and the possible relation to equilibrium thermodynamics. A global incoherent dynamic structure factor, \sqw, dominated by hydrogen, is observed in neutron scattering measurements. A first consideration is that the hydrogen atoms in a protein occupy a spectrum of sites and the fluctuation of H in these sites follows a wide distribution, which can be modelled using a Weibull form.\cite{Meinhold:08} This heterogeneity in the distribution of mean-square displacements leads to dynamic structure factors, \sqw, that deviate from Gaussian behaviour in the scattering wave vector $Q$. The heterogeneity introduces a correction to fourth order in $Q$ that can be used to extract the variance of the distribution of mean-square displacements.\cite{Becker:03,Yi:12}

Similarly, the global MSD obtained from neutron scattering measurements depends on the energy resolution employed. 
The MSD is obtained from the elastic (\om~ = 0) component of 
the resolution broadened dynamic structure factor (DSF), \srqwe, as 
\begin{equation} \label{e1}
\langle r^2\rangle _{R} = -3\frac{d \ln S_R(\textbf{Q},\omega =0)}{dQ^2}. 
\end{equation}
The \rsslope~ extracted in this way is the MSD after it has had time to develop 
over a limited time, 0 $< t <$ \tr~ only. The time \tr~ is set by the width of the energy resolution of the 
neutron instrument, \tr~$\simeq \hbar/W$ where $W$ is the full width at half maximum (FWHM) 
of the resolution function. Typical instrument energy resolutions lie in the range 100 $\mu$eV $> W >$ 1$
\mu$eV which correspond to evolution times 10 ps $<$ \tr $<$ 1 ns. Over this time 
range the extracted \rsslope~ is still increasing with decreasing $W$ (increasing \tr) indicating that the intrinsic, long time (\tr $\rightarrow \infty$) value of \rs~ has not been observed.\cite{Nakagawa:10,Jasnin:10,Wood:08,Daniel:03}

In an earlier paper\cite{Vural:12}, we proposed a method to extract the intrinsic, long time MSD \rs~ from resolution dependent data. In the method, the intrinsic \rs~ was defined in terms of the t = $\infty$ limit of the incoherent intermediate scattering function (ISF), \iqt, as 
\begin{equation} \label{e1a}
I_{\infty} = I(\textbf{Q},t = \infty)= \exp(- \frac{1}{3}Q^2 \langle r^2 \rangle).
\end{equation}
The method consists of constructing a model \iqt~ that includes \ii~ and \rs, calculating the corresponding \srqwe~ including the resolution width, $W$, and fitting the model to the observed \srqwe. The intrinsic \rs~ is obtained from the fit as a fitting parameter. In this way an intrinsic, long time \rs~ was obtained from resolution dependent data.

The incoherent intermediate scattering function, \Iiqt, that is observed in neutron measurements can also be calculated directly from simulations.\cite{Smith:91} A comparison between the simulation-derived and observed MSD can be made by calculating the simulation-derived MSD in exactly the same way as it is obtained from experiment.\cite{Hayward:03,Tarek:00a,Roh:06,Wood:08} That is, from the simulated \Iiqt, the incoherent DSF \sqw~ is calculated including the instrument resolution width, $W$. The MSD is obtained from the slope of the calculated resolution broadened \srqw, at small Q using Eq. (\ref{e1}). This effectively compares the MSD after it has evolved for a specific time $\tau$ set by $W$. In this way excellent agreement between simulated and 
observed MSD has been achieved. This also opens the question: could simulated intrinsic, long time \rs~ be obtained by fitting a model to finite time values of a calculated \Iiqt?  

To explicitly recognize the time dependence of \rsr, the concept of neutron time windows was introduced, a concept in which the effects arising from the finite energy resolution are fully integrated.\cite{Becker:03,Becker:04} The concept has also been incorporated into the formalism that describes the dynamics accompanying the glass transition in molecular systems. When the protein relaxation time decreases with temperature, as it usually does, it has been shown that measurement of \rsr~ over a finite time window can introduce an apparent DT when there is no actual change in the elastic incoherent DSF, \sqwe.\cite{Becker:03,Becker:04} To avoid this issue, identification and use of an intrinsic, long time MSD \rs~ would be helpful.

Several MD studies have been performed aimed at understanding the origin of elastic neutron scattering from proteins.\cite{Hayward:02,Hayward:03,Hayward:03a,Meinhold:08,Yi:12,Miao:12,Becker:03,Becker:04} These studies have characterized in detail the contributions of time dependence, of non-Gaussian behavior and of dynamical heterogeneity to elastic scattering on the ps-ns timescales. The strong time-dependence of \rsr~ revealed in these studies further opens the question as to whether an intrinsic long-time, time-independent MSD can be obtained from simulations. Given that folded proteins have well-defined three-dimensional structures, a well defined long time, intrinsic MSD would seem to be reasonable and should exist. That is, the internal mean-square displacement arising from internal motions in proteins should converge to a plateau as a function of time and a parallel in proteins of the Debye-Waller and X-ray B factors found in the crystalline state should exist. Given this assumption, the question focuses on developing a method to extract the intrinsic \rs~ from finite time simulations. 

In addition to the ISF \Iiqt, an MSD can be calculated directly from simulations. This MSD is
\begin{equation} \label{e2}
\Delta ^2 (t) = \langle [r(t)-r(0)]^2\rangle \equiv \frac{1}{N}\sum _{i=1}^N \langle 
[r_i(t)-r_i(0)]^2\rangle,
\end{equation}
where $r_i(t)$ is the position of nucleus $i$ in the protein at time $t$.  After long times $t 
\rightarrow \infty$, when the positions $r_i(t)$ and $r_i(0)$ are no longer correlated ($\langle 
r(\infty)r(0)\rangle = 0$), the $\Delta ^2 (t)$ reduces to  
\begin{equation} \label{e3}
\Delta ^2 (t \rightarrow \infty) = \langle r^2(\infty)\rangle + \langle r^2(0)\rangle = 
2\langle r^2\rangle_{MD}.
\end{equation}
In this way an \rsmd~ = $\Delta ^2 (t \rightarrow \infty)/2$ for H in proteins can be defined and calculated. However, we emphasize that this \rsmd~ is not the same as the intrinsic \rs~ defined in terms of \iqt~ in Eq. (\ref{e1a}). Firstly, the average over the nuclei in the protein made when the full \Iiqt~ is represented by a global \iqt~ is not the same as the average over the nuclei made in Eq. (\ref{e2}). Also the $\Delta^2(t)$ does not converge to a constant, infinite time value within accessible simulation times of 1 to 100 ns.\cite {Hayward:02,Hamon:06,Calandrini:08} Thus the correlations $\langle r(t)r(0)\rangle$ have not vanished and Eq. (\ref{e3}) is not obviously valid within accessible simulation times. Essentially, a \rsmd = $\Delta ^2 (t \rightarrow \infty)/2$ cannot be calculated within currently accessible simulation times. Keeping in mind these differences, \dsq /2 is a physically interesting, time dependent quantity to calculate. It is especially useful in determining the time scales needed for \dsq/2 to converge toward a fixed value. However, even the converged value may differ from \rs.  

In this context, the goal of the present paper is to propose a method for 
obtaining an intrinsic, long time, $t \rightarrow \infty$, value of the MSD from finite-time simulations.
We seek an intrinsic MSD from simulation that is defined exactly as in neutron scattering measurements, i.e. in terms of the global \iqt~ in Eq. (\ref{e1a}). The procedure is to construct a model of the global \iqt~ which contains \rs~ and fit the model to finite time simulations of $I_{inc}(\textbf{Q},t)$. Explicitly, the $I_{inc}(\textbf{Q},t)$ observed in neutron scattering experiments and calculated from simulations is \cite{Rog:03},
\begin{eqnarray} \label{e4}
I_{inc}(\textbf{Q},t) &=& \frac{1}{N}\sum_{i=1}^N b_i^2 \langle \mathrm{e}^{-i\textbf{Q}\cdot\textbf{r}_i(t)}\mathrm{e}^{i\textbf{Q}\cdot\textbf{r}_i(0)}\rangle.
\end{eqnarray} 
In Eq. (\ref{e4}), $b_i$ is the incoherent scattering length of nucleus $i$ in the protein. As indicated above, the $b_i$ of hydrogen is more than 20 times larger than the $b_i$ of other nuclei typically found in proteins. For this reason the scattering from hydrogen, which is also 
almost entirely incoherent, dominates ISF. In the analysis of neutron scattering 
experiments it is usual to represent the ISF in Eq. (\ref{e4}) summed over all nuclei by a 
global $I(\textbf{Q},t)$ which represents the whole protein,

\begin{equation} \label{e5}
 I(\textbf{Q},t) = \langle \exp(-i\textbf{Q}.r(t)) \exp(i\textbf{Q}.r(0))\rangle .
\end{equation}
Following the same procedure we construct a model of \iqt. The model \iqt~ is separated into a time 
independent part, $I(\textbf{Q},t = \infty)$,  and a time dependent part $I'(\textbf{Q},t)$. The time independent part, 
in the Gaussian approximation, is \ii~=  $I(\textbf{Q},t = \infty)$ given by Eq. (\ref{e1a}).
$I_{\infty}$ is the familiar Debye-Waller factor. We define the intrinsic MSD as the \rs~ that appears in $I_{\infty}$. 

To implement the method, we first calculate \Iiqt~ from MD simulations of a hydrated protein (lysozyme) using Eq. (\ref{e4}). We then fit the model of the global \iqt~ in Eq. (\ref{e5}), which contains \ii~ and \rs, to the simulated \Iiqt. We treat the \rs~ in the model as a free fitting parameter. In this way we obtain an infinite time value of the MSD $\langle r^2\rangle$ from fits to simulation data at finite $t$.

We test the present method using simulations of lysozyme at several temperatures and of two 
simulation lengths, $t = 100$ ns and $1$ $\mu$s. From the simulations, we calculate $I_{inc}(\textbf{Q},t)$ given by 
Eq. (\ref{e4}) out to $1$ ns and $10$ ns, respectively. From fits of the model \iqt~ to the calculated 
$I_{inc}(\textbf{Q},t)$, we obtain corresponding fitted values of $\langle r^2\rangle$. We find that $\langle r^2\rangle$ is the same 
for the two simulation times, consistent with $\langle r^2\rangle$ representing a time-independent, long 
time MSD. The intrinsic \rs~ is approximately twice \rsslope, the MSD calculated for motions out to 1.5 ns. 
A plot of $\langle r^2\rangle$ versus temperature shows a break in slope at 140 K and a second dynamical transition at $T_D = 220$ K, as has been observed and calculated for proteins experimentally. According to this model, then, the dynamical transitions are intrinsic properties of proteins. While the appearance of the DT and transition temperature, \td, may be modified by experimental time windows, the transitions exist independently of  finite experimental time windows.

\section{Molecular Dynamics Simulation}

\begin{figure}[h]
\includegraphics[scale=0.85,angle=0]{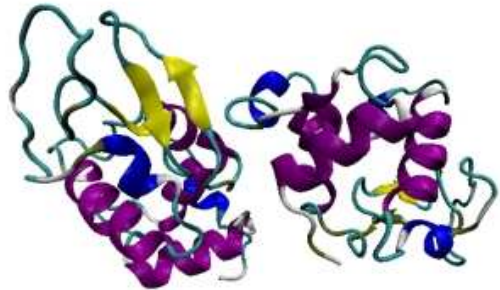}
\vspace{0.5cm}
\caption{Two lysozyme molecules of random relative orientation selected by GROMACS.}
\label{fig:f1}
\end{figure}

Two lysozyme molecules (1AKI\cite{Artymiuk:82}) were arbitrarily oriented as shown in Fig.\ref{fig:f1} and placed in a simulation box of dimensions $6.5$ nm $\times 3.4$ nm $\times 3.6$ nm. The lysozyme molecules inside the simulation box were surrounded by 636 water molecules, corresponding to the hydration level $h=0.4$ g water/g protein. The box was replicated using periodic boundary conditions to mimic the environment of an experimental powder sample.  Similar simulation systems are discussed in the literature.\cite{Hong:11,Hong:12,Lagi:08,Oleinikova:05,Dirama:05,Tarek:00}

The system was simulated using GROMACS 4.5.1.\cite{Hess:08} The OPLS-AA force field\cite{Jorgensen:88} was used for the protein and the TIP4P force field\cite{Horn:04} for the water. The van der Waals interaction was truncated at $1.4$ nm, and the electrostatic interaction was represented using the Particle Mesh Ewald method\cite{Essmann:95} with a real-space cutoff of $0.9$ nm. All bonds including hydrogen bonds were constrained with a linear constraints solver algorithm (LINCS).\cite{Hess:97} The energy of the system was first minimized using 50000 steepest descent steps. The system was then  equilibrated in the NVT (mole-volume-temperature) ensemble at each temperature investigated for 10 ns and in the NPT (mole-pressure-temperature) ensemble at 1 bar for 10 ns. The Nose-Hoover algorithm\cite{Hoover:85} with a coupling time $\tau =1$ ps and the Parrinello-Rahman algorithm\cite{Parrinello:81} with a coupling time $\tau =3$ ps were used for the temperature coupling and pressure coupling, respectively. 

Simulations of 100 ns length were performed at 18 different temperatures between $80$ K and $300$ K. Simulations of $1$ $\mu$s were made at 5 temperatures, at $100$ K and then in steps of $50$ K to $300$ K. The data were collected every $10$ ps at each temperature for both simulations.

\section{The ISF and Model $I(\textbf{Q},t)$}

\begin{figure*}
\includegraphics[scale=0.26,angle=0]{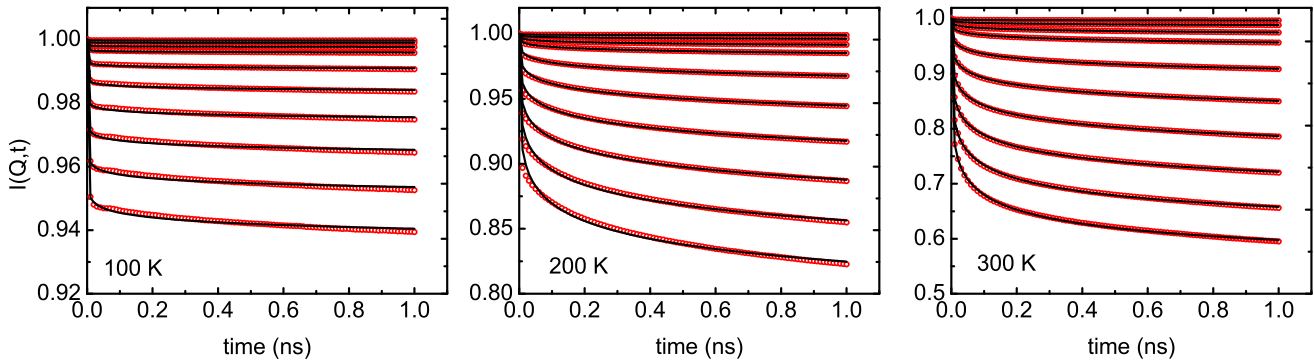}
\vspace{0.5cm}
\caption{The intermediate scattering function (ISF), \Iiqt, for $0 < t < 1$ ns of hydrated lysozyme (h = 0.4) obtained from a $100$ ns MD simulation (open red circles) and fits of the model ISF \iqt~ in Eq. (\ref{e11}) (blue solid lines) to the \Iiqt~ at $100$ K, $200$ K and $300$ K. From top to bottom, $Q$: $0.1$, $0.2$, $0.3$, $0.4$, $0.6$, $0.8$, $1$, $1.2$, $1.4$ and $1.6$ \AA $^{-1}$.}
\label{fig:f2}
\end{figure*} 

\begin{figure}[h]
\includegraphics[scale=0.3,angle=0]{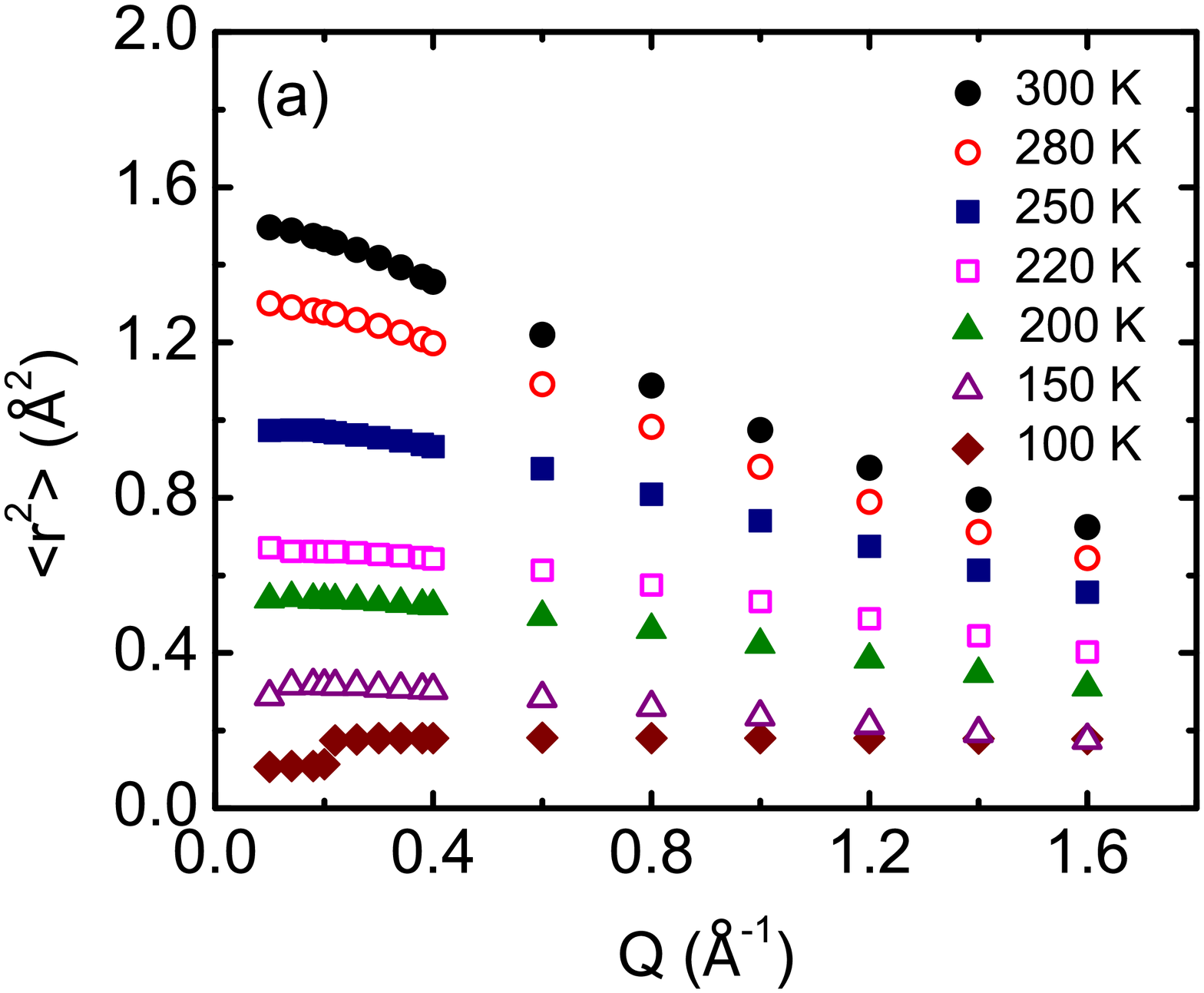}\\
\includegraphics[scale=0.3,angle=0]{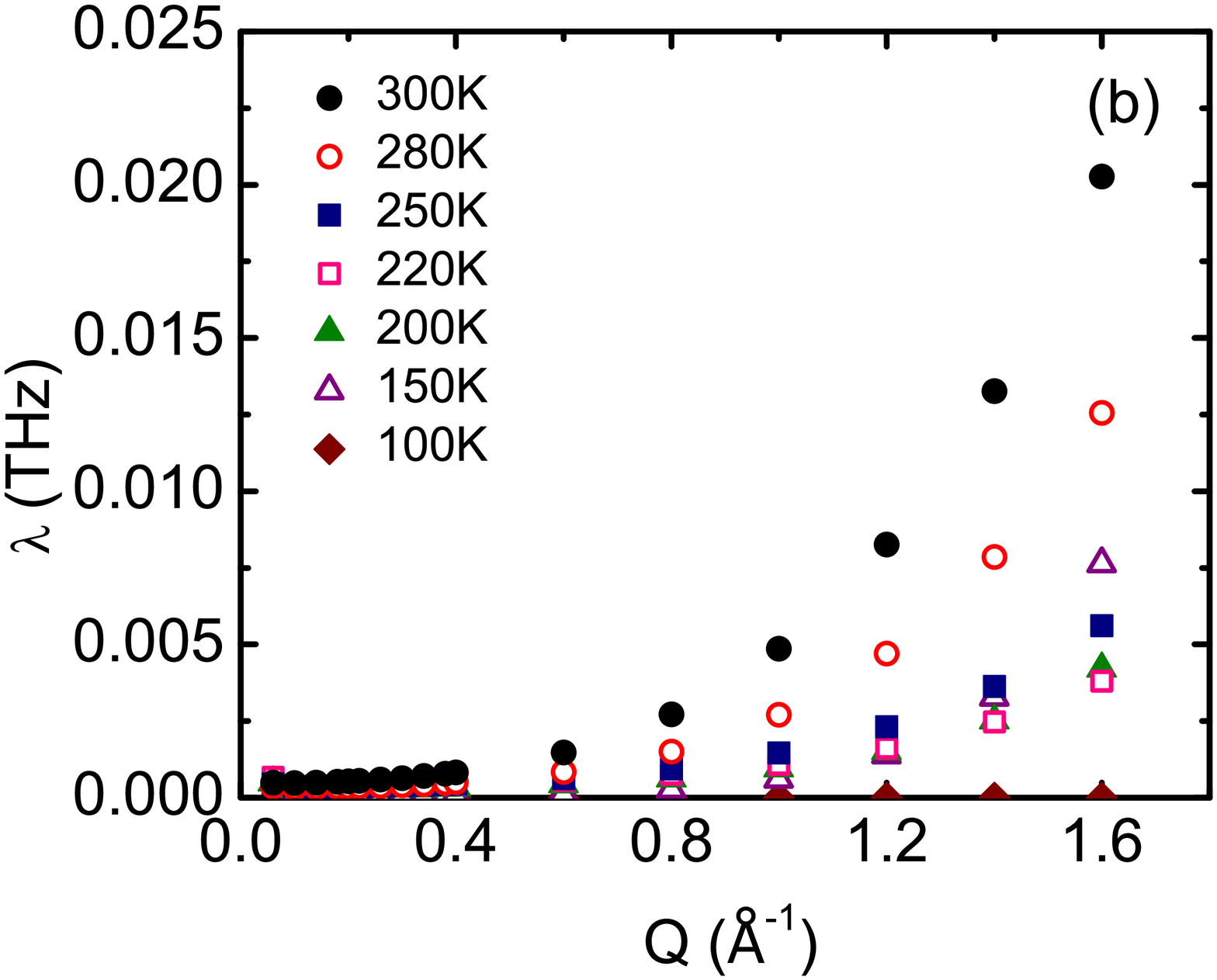}
\vspace{0.5cm}
\caption{Parameters of the model ISF \iqt~ of Eq. (\ref{e11}) obtained from fits of the model to the simulations shown in Fig. \ref{fig:f2}: (a) The intrinsic MSD, \rs~ and (b) the relaxation parameter, $\lambda$, versus $Q$ at temperatures $100$ K to $300$ K. }
\label{fig:f3}
\end{figure}

\begin{figure}
\includegraphics[scale=0.3,angle=0]{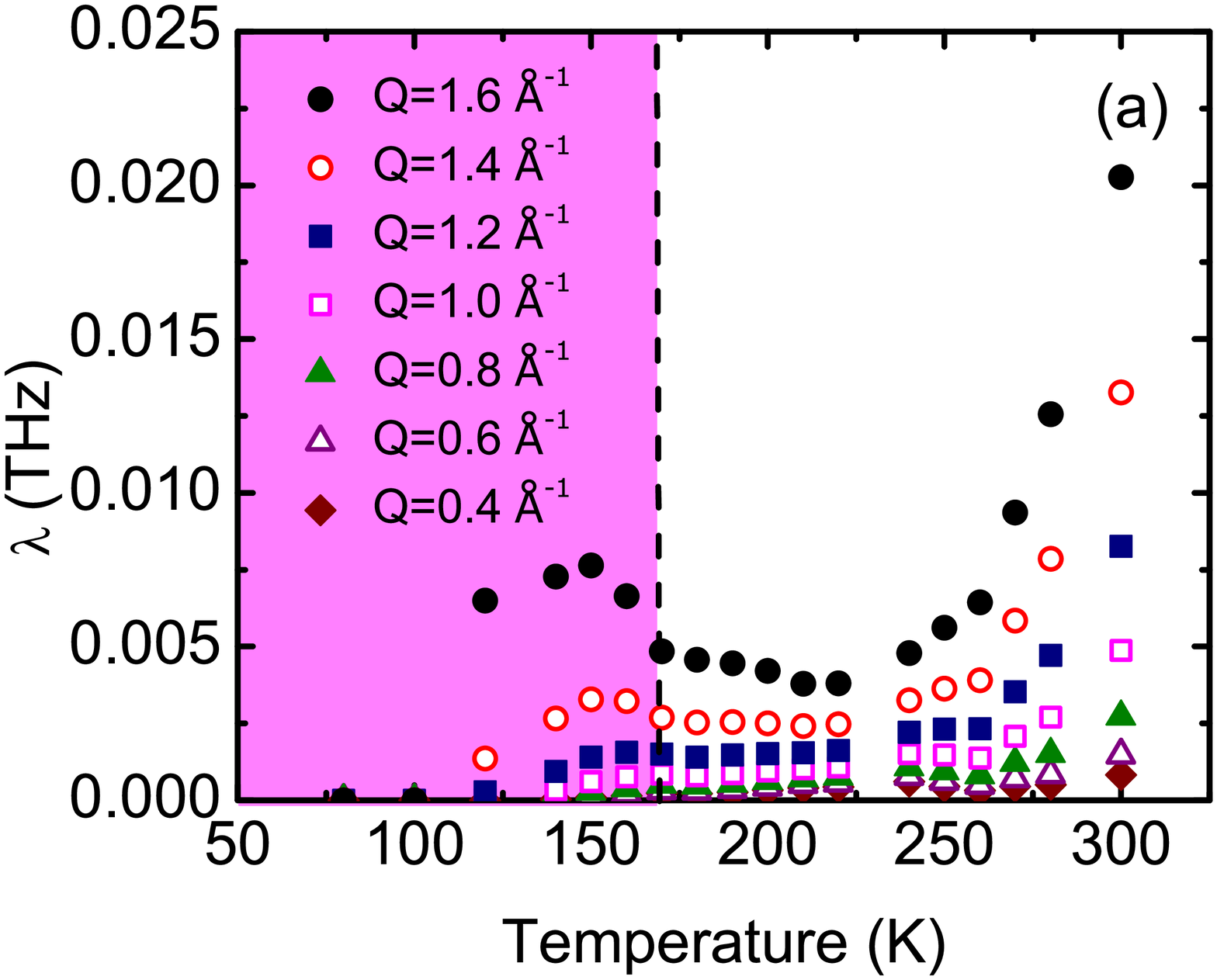}\\
\includegraphics[scale=0.3,angle=0]{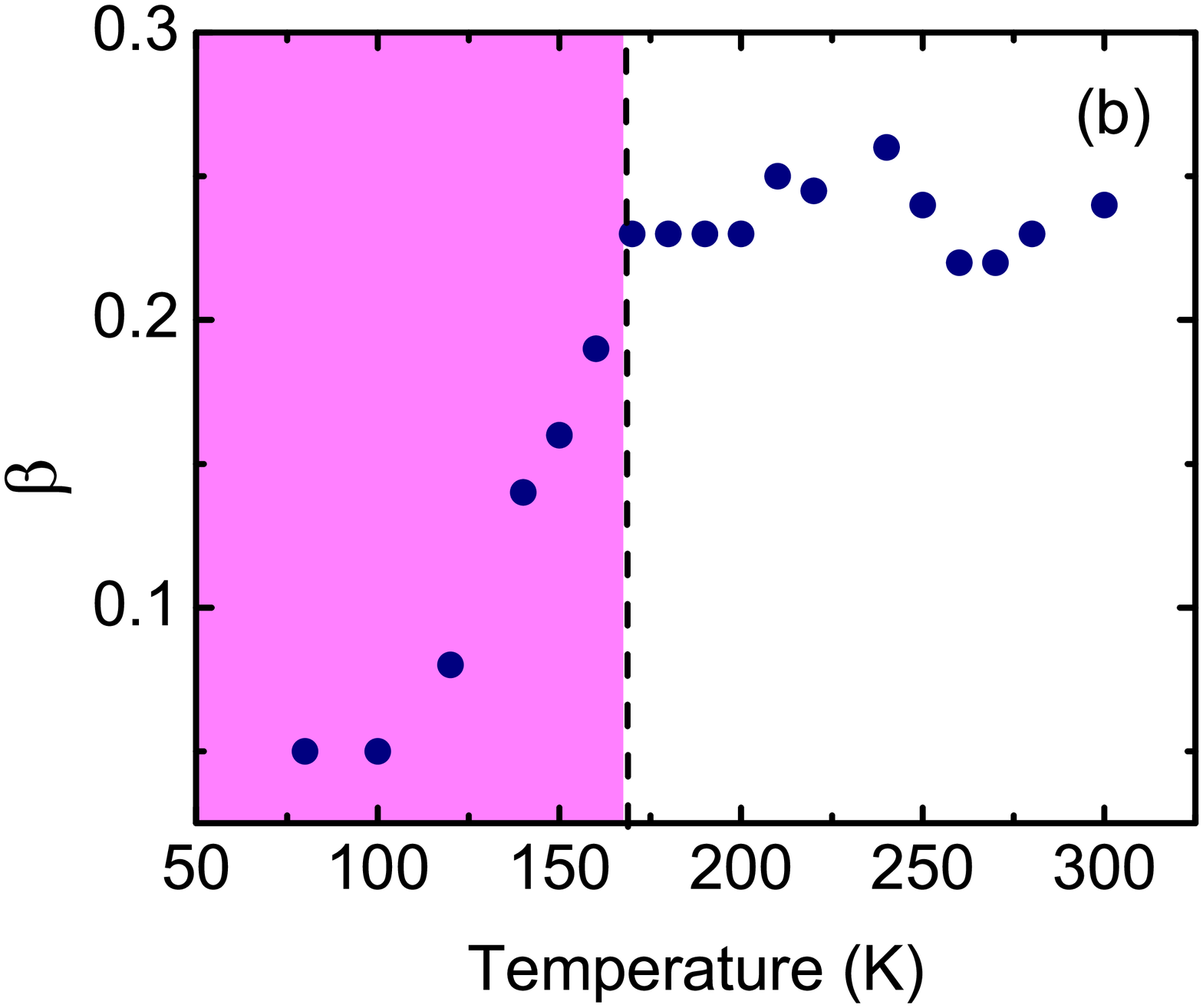}
\vspace{0.5cm}
\caption{As Fig. \ref{fig:f3} for (a) the relaxation parameter $\lambda$ and (b) the stretched exponential parameter $\beta$ versus temperature. From bottom to top, $Q$: $0.4$, $0.6$, $0.8$, $1$, $1.2$, $1.4$ and $1.6$ \AA $^{-1}$.}
\label{fig:f4}
\end{figure}

\begin{figure}
\includegraphics[scale=0.3,angle=0]{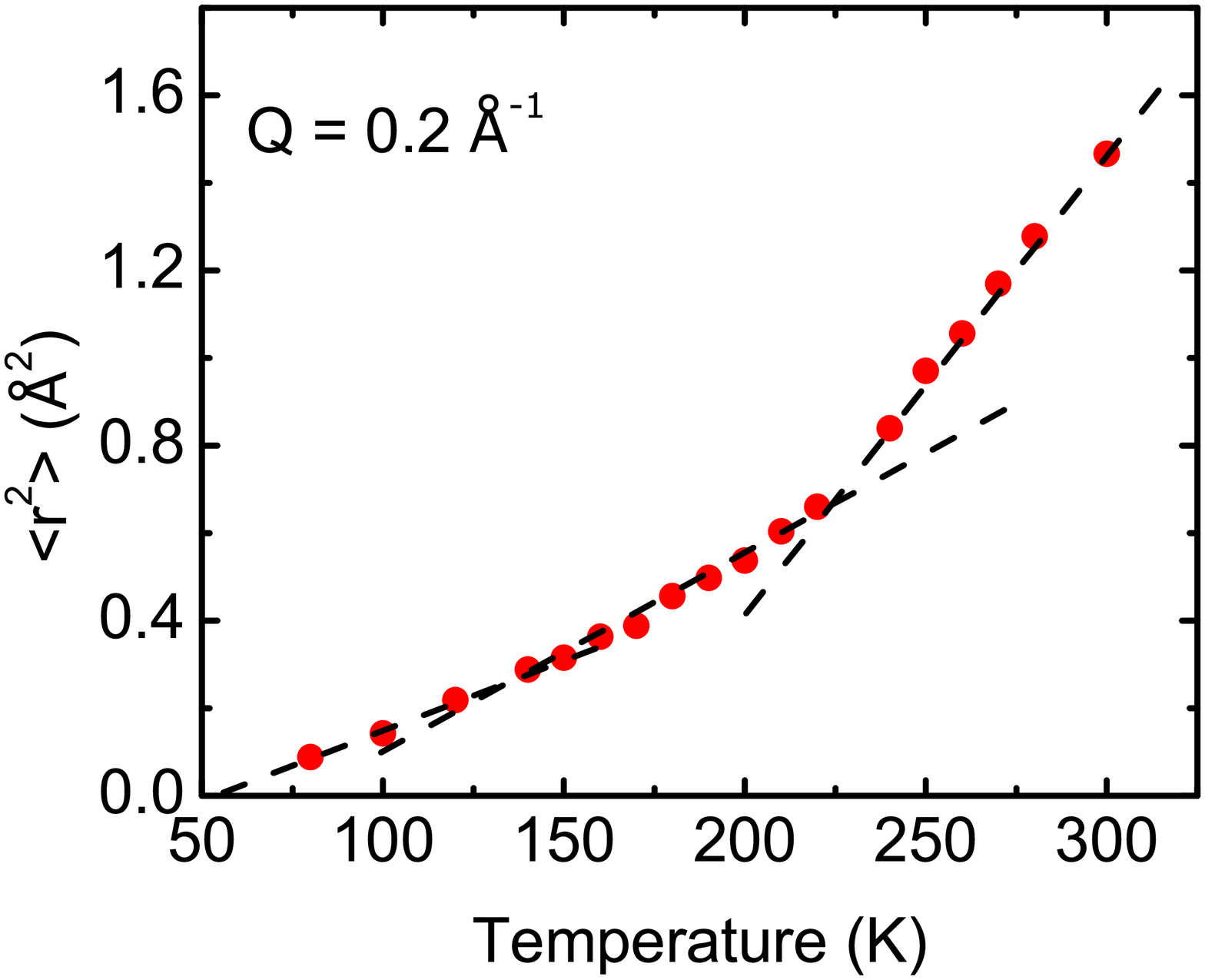}
\vspace{0.5cm}
\caption{ The intrinsic MSD, \rs, defined in Eq. (\ref{e1a}), obtained from fits of the model ISF, Eq. (\ref{e11}) to simulations of \Iiqt~ at $Q = 0.2$ \AA $^{-1}$. The intrinsic \rs~ shows a break in slope at $T \simeq~140$ K and $T \simeq~220$ K.}
\label{fig:f5}
\end{figure}

To obtain the intrinsic MSD, \rs, following the procedure outlined in the Introduction we firstly calculated the intermediate incoherent scattering function, \Iiqt, defined in Eq. (\ref{e4}). The positions $r_i(t)$ of each nucleus $i$ in the protein were generated in the two MD simulations described above, one of length 100 ns and the other 1 $\mu$s. Using the $r_i(t)$, the \Iiqt~ is calculated directly for times out to 100 ns and 1 $\mu$s, respectively. To improve statistics, each simulation was divided into at least 100 segments and \Iiqt~ recalculated as an average over these segments. In this way \Iiqt~ was calculated to 1 ns from the 100 ns MD data and to 10 ns from the 1 $\mu$s simulation.

Next we developed a model for the global $I(\textbf{Q},t)$ defined in Eq. (\ref{e5}) which contains the intrinsic MSD, \rs, defined in Eq. (\ref{e1a}). The model is obtained firstly by separating $I(\textbf{Q},t)$ into a time independent, $t = \infty$, part ($I_{\infty} = I(\textbf{Q},t = \infty)$) and a time dependent part  $(I(\textbf{Q},t)-I_{\infty})$ part,
\begin{equation} \label{e7}
I(\textbf{Q},t) = I_{\infty} + (I(\textbf{Q},t)-I_{\infty}),
\end{equation}
From Eq.~(\ref{e5}), 
\begin{eqnarray} \label{e8}
I_{\infty} = I(\textbf{Q},t = \infty) &=& \langle \exp(-i\textbf{Q}.r(\infty)) \exp(i\textbf{Q}.r(0))\rangle \nonumber \\
   &=& \exp(- \frac{1}{3}Q^2 \langle r^2 \rangle + O(Q^4))
\end{eqnarray}
is the infinite time limit. To obtain the last expression we assume: (1) that $r(\infty)$ and $r(0)$ are 
completely uncorrelated so that the averages of them are independent, (2) that the system is 
translationally invariant in time (no CM motion) so that  $r(\infty) = r(0)$ and (3) that in a cumulant 
expansion of $\langle \exp (-i\textbf{Q}.r)\rangle$, cumulants beyond the second are negligible. The 
last assumption is valid if $Q$ is small or if the distribution over $r$ is approximately a Gaussian distribution. 
The cumulants beyond the second vanish exactly for all $Q$ if the distribution over $r$ is exactly 
Gaussian.

The time dependent part of \iqt~ has the limits
 \begin{displaymath}
   I'(\textbf{Q},t) = I(\textbf{Q},t)-I_{\infty}  = \left\{
     \begin{array}{lr}
       1-I_{\infty} & t=0 \\
       0   & t = \infty 
     \end{array}
   \right .
\end{displaymath} 
We model this by the function
\begin{equation} \label{e9}
I'(\textbf{Q},t) = (1-I_{\infty}) C(t),
\end{equation}
where $C(t)$ has the limits $C(t=0)=1$, $C(t=\infty)=0$. An example is the stretched exponential function, 
\begin{equation} \label{e10}
C(t) = \exp (-(\lambda t)^{\beta}),
\end{equation}
where $\lambda$ and $\beta$ are constants. The $C(t)$ represents the decay of correlations in the protein. Collecting, the model is

\begin{equation} \label{e11}
I(\textbf{Q},t) = I_{\infty}(\textbf{Q}) + (1- I_{\infty}(\textbf{Q}))C(t),
\end{equation}
which is constructed to have the correct limits at $t=0$ and $t=\infty$ and to have a plausible 
representation of several motional decay processes at intermediate times described by $C(t)$. We fit the model $I(\textbf{Q},t)$ in Eq. (\ref{e11}) to the calculated \Iiqt~ to determine \rs, $\lambda$~ and $\beta$. The model $I(\textbf{Q},t)$ is 
the same as we used previously\cite{Vural:12} to fit neutron data except that $C(t)$
is a stretched exponential rather than a simple exponential used previously. In fits to neutron data we found\cite{Vural:12} that the data was not sufficiently precise to distinguish between a stretched and simple exponential. In contrast, a simulation-derived \Iiqt~ is more discriminating. 

\begin{figure*}
\includegraphics[scale=0.26,angle=0]{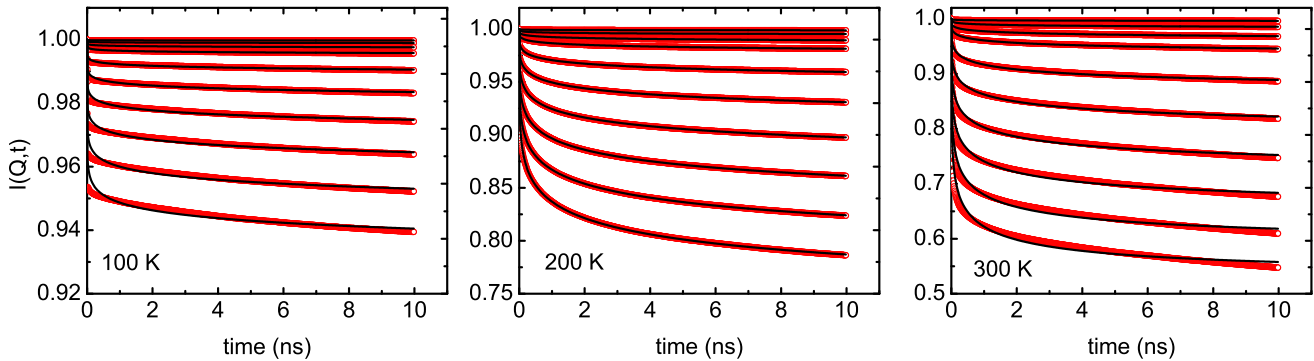}
\vspace{0.5cm}
\caption{The calculated \Iiqt~, for $0 < t < 10$ ns of hydrated lysozyme obtained from a $1$ $\mu$s MD simulation (open circles) and fits of the model ISF \iqt~ in Eq. (\ref{e11}) to the data (solid lines) at $100$ K, $200$ K and $300$ K. From top to bottom, $Q$: $0.1$, $0.2$, $0.3$, $0.4$, $0.6$, $0.8$, $1$, $1.2$, $1.4$ and $1.6$ \AA $^{-1}$.}
\label{fig:f6}
\end{figure*} 

\section{Results}

In this section, we present the fits of the model \iqt~ given by Eq. (\ref{e11}) to the calculated ISF data for lysozyme. The model of \iqt~ includes three fitting parameters: the intrinsic MSD, \rs, defined in Eq. (\ref{e1a}), the relaxation parameter $\lambda$ and the stretched exponential parameter $\beta$  defined in Eq. (\ref{e10}). The goal is to determine the intrinsic MSD \rs~ of H in lysozyme and to obtain values for the relaxation parameters $\lambda$ and $\beta$. 

\subsection{Intrinsic MSD} 

\subsubsection{$100$ ns MD simulation }

Fig.~\ref{fig:f2} shows the intermediate scattering function, \Iiqt, for times $0 < t < 1$ ns calculated from Eq. (\ref{e4}) using the $100$ ns simulation data. Although \Iiqt~ was calculated at 18 temperatures, only 3 temperatures are shown in Fig.~\ref{fig:f2}. The solid lines are fits of the model \iqt~ to the calculated \Iiqt. The fits are better at higher temperatures than at lower temperatures. Particularly, at temperatures below approximately $170$ K, the parameters $\lambda$ and $\beta$ that appear in the relaxation function $C(t)$ are not well determined as we discuss below. 

Figs.~\ref{fig:f3} and \ref{fig:f4} show the best fit values of the fitting parameters \rs, $\lambda$ and $\beta$. From Fig.~\ref{fig:f3}a, we see that $\langle r^2\rangle$ is $Q$ dependent and is larger and approximately independent of $Q$ at low $Q$. This $Q$ dependence is similar to that obtained from fits to observed data  (to \sqwe). The \rs~ is determined chiefly by the value of \iqt~ at long $t$, i.e. by how far $I(Q, \infty)$ lies below $I(Q,t = 0) = 1$.

From Fig.~\ref{fig:f3}b, we see that $\lambda$ is also $Q$ dependent, with $\lambda \varpropto Q^2$ approximately, as found in other simulations.\cite{Tarek:00,Chen:10}  We found $\beta$ only weakly dependent on $Q$ and we used an average over several $Q$ values with some adjustments to obtain smooth behavior as a function of temperature. The resulting temperature dependence of $\beta$ is shown in Fig.~\ref{fig:f4}b. The $\lambda$ and $\beta$ are not well determined at temperatures below approximately $170$ K. Essentially, at low temperature \Iiqt~ decreases rapidly over a short time $t$ and thereafter changes slowly. This time dependence is consistent with harmonic motion as shown in Smith \textit{et al.}\cite{Smith:86,Smith:90} and can be approximately reproduced by a range of $\lambda$ and $\beta$ values. The \rs~ remains well determined at low temperature since it is determined chiefly by \Iiqt~ at long times.

The decrease of the best fit value of \rs~ with increasing $Q$ can have two origins. Firstly, at low $Q$ we are sampling longer range phenomena. Long range motions could contribute fully to $\langle r^2\rangle$ at low $Q$ whereas they could be limited or cut off at high $Q$ leading to a smaller or limited $\langle r^2\rangle$ at high $Q$. Since this is a real physical effect, values of \rs~ obtained from data (e.g. \sqwe) at low $Q$ ($0 < Q < 0.4$ \AA$^{-1}$) are usually selected. Secondly and most importantly, Zheng \textit{et al.}\cite{Yi:12} have shown that \iqt~ departs from a Gaussian approximation because of dynamical heterogeneity. The heterogeneity introduces a $Q^4$ term in Eq. (\ref{e8}) which becomes significant at larger $Q$. This means that \rs~ obtained from data at small $Q$ must be selected. 

Fig.~\ref{fig:f5} shows the intrinsic MSD \rs~ obtained from fits to \Iiqt~ at $Q = 0.2$ \AA$^{-1}$. The \rs~ versus temperature shows a clear break in slope at around $T = 140$ K. This has been seen previously in simulations and arises from the activation of the dynamics of hydrophobic groups, i.e. the onset of proline puckering and the rotation of methyl groups at around $140$ K.\cite{Roh:06,Miao:12} A break in slope of \rs~ versus $T$ near $140$ K has also been observed in several proteins. A second break in slope is seen at the well documented dynamical transition (DT), at $T_D \simeq 220$ K, associated with the onset of new larger amplitude motions of hydrophilic groups in which the hydration water plays a determining role. The intrinsic, long time \rs~ shows the onset of both hydrophobic and hydrophilic (DT) motions.    

\subsubsection{$1$ $\mu$s MD simulation } 

We turn now to the \Iiqt~ calculated from the $r_i(t)$ generated in the 1 $\mu$s simulations. Five temperatures from $100$ K to $300$ K were simulated. As before, the simulation data was divided into at least 100 segments (time slices), each spanning a time $t$, $0 < t < 10$ ns, and \Iiqt~ was calculated as an average over the time slices. The resulting \iqt~ are shown in Fig.~\ref{fig:f6} at three temperatures. The solid line in Fig.~\ref{fig:f6} is again a fit of the model \iqt~ given by  Eq. (\ref{e11}) with \rs, $\lambda$ and $\beta$ as treated as free fitting parameters. As in Fig.~\ref{fig:f6}, the fits are better at the higher temperatures, although the fit is better at $200$ K than at $300$ K for times out to $10$ ns. 

\begin{figure}[h!]
\includegraphics[scale=0.288,angle=0]{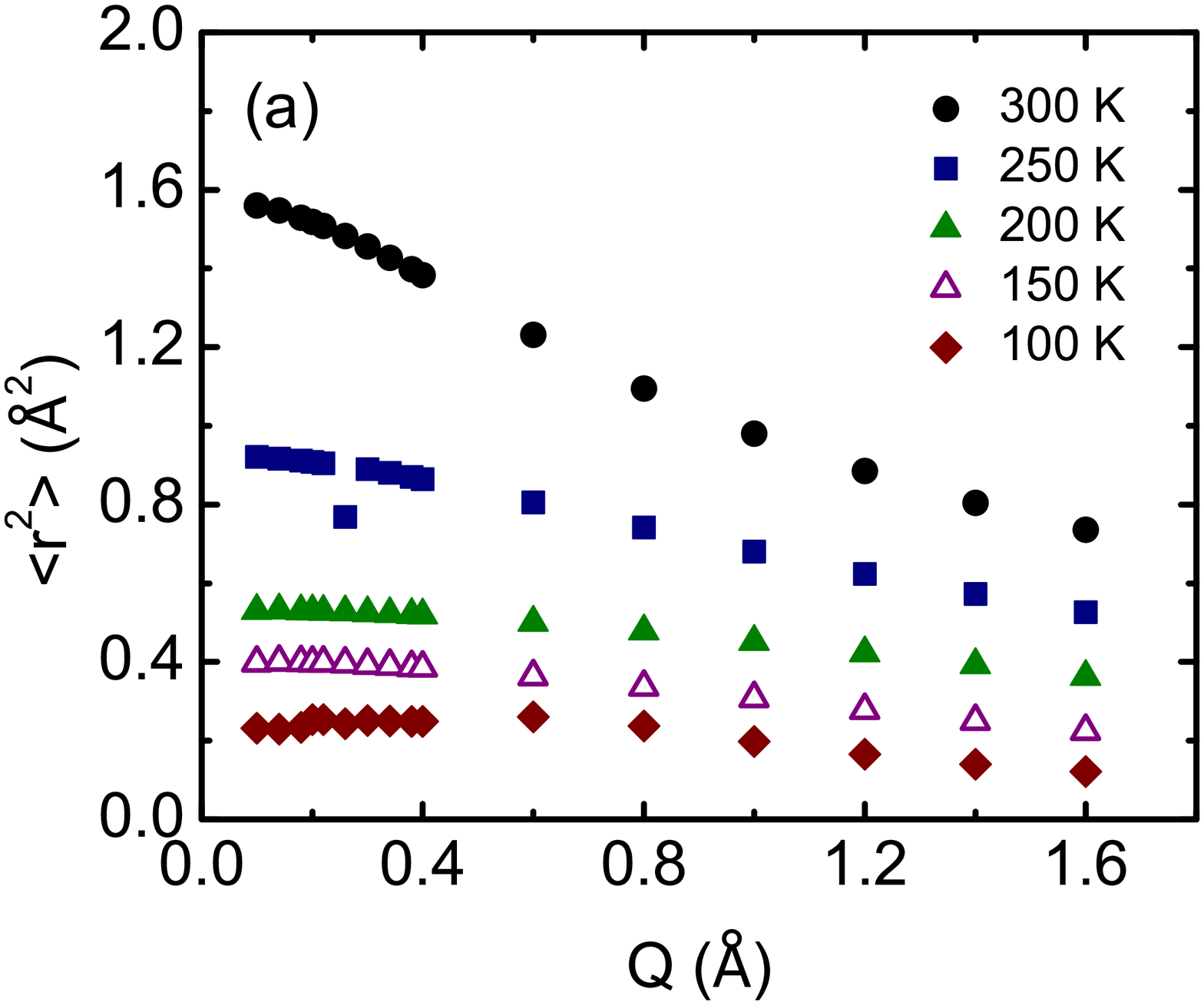}\\
\includegraphics[scale=0.288,angle=0]{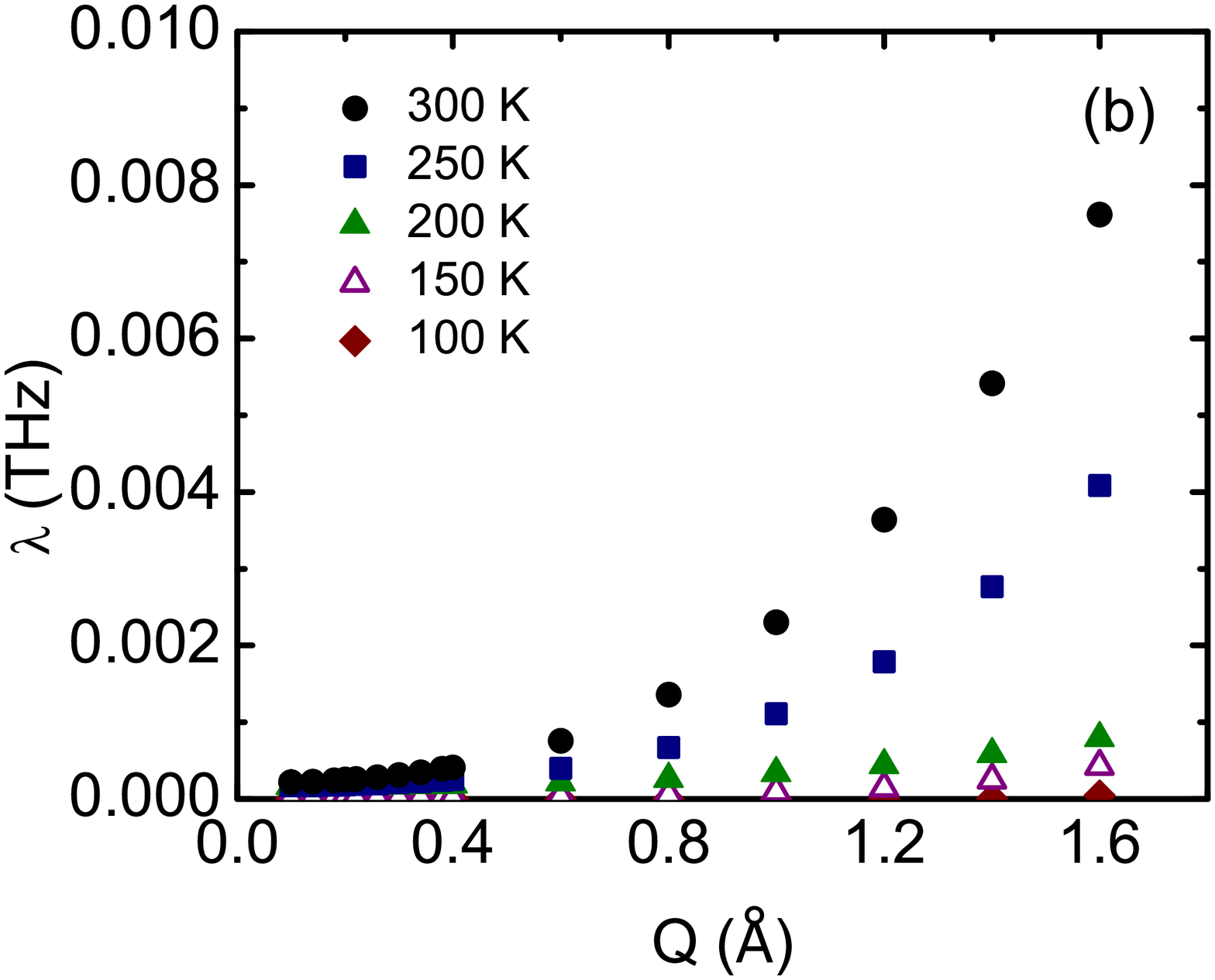}\\
\includegraphics[scale=0.288,angle=0]{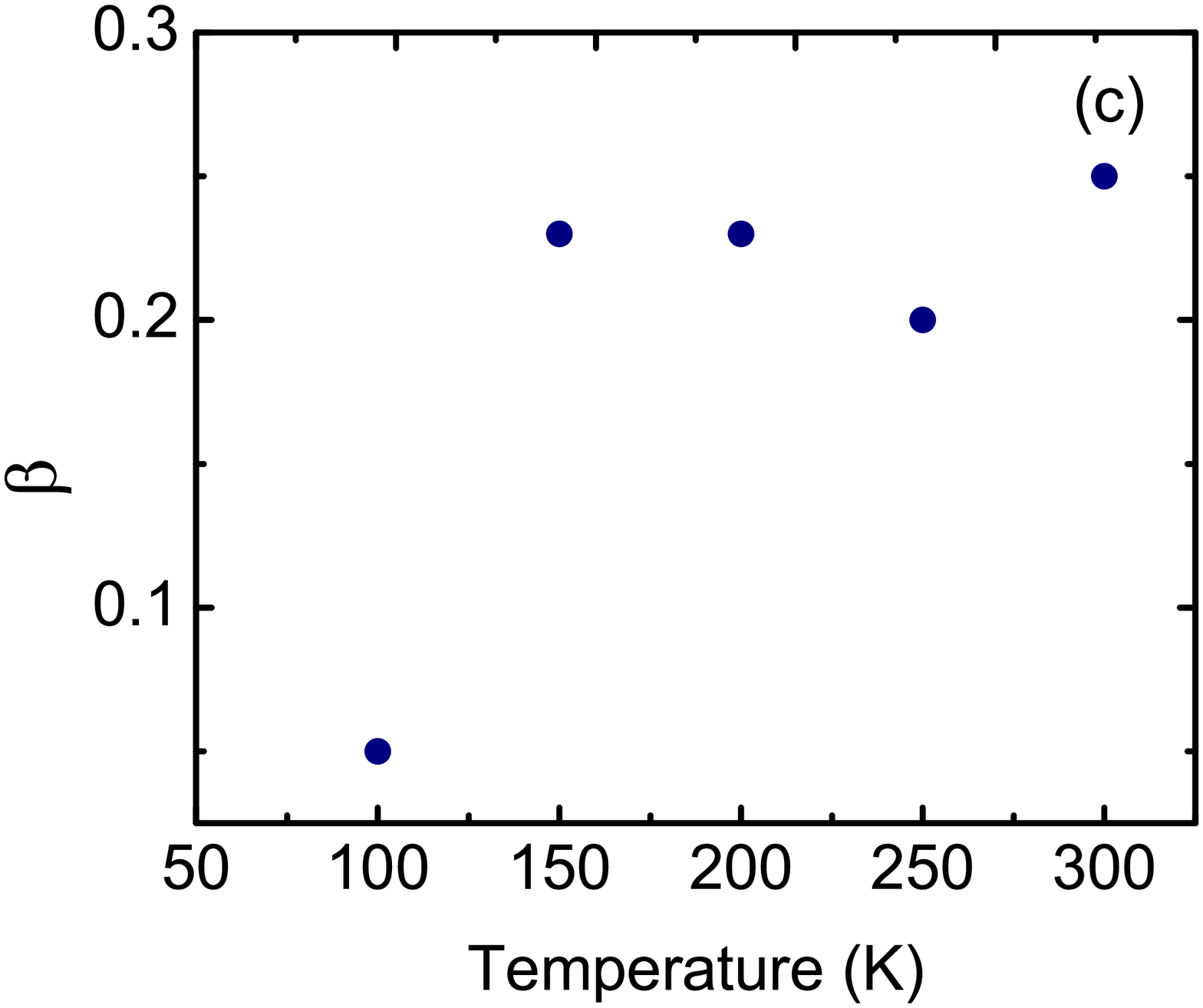}
\vspace{0.5cm}
\caption{Parameters of the model ISF, Eq. (\ref{e11}), obtained from the fits of \iqt~ to the calculated \Iiqt, for $0 < t < 10$ ns shown in Fig. (\ref{fig:f6}): (a) the intrinsic MSD, \rs, (b) the relaxation parameter, $\lambda$, and (c) the stretched exponential parameter, $\beta$, at five temperatures: $100$ K, $150$ K, $200$ K, $250$ K and $300$ K.}
\label{fig:f7}
\end{figure}

The best fit values of the intrinsic MSD \rs, $\lambda$ and $\beta$ are shown in Fig.~\ref{fig:f7}. The \rs~ decreases with increasing $Q$ as was found for the shorter simulation. The absolute values of \rs~ are also consistent with those obtained from the shorter simulation expect, possibly, at $100$ K. $\lambda ^2$ is approximately proportional to $Q^2$ as found in the shorter simulation. The absolute values of $\lambda$ obtained from fits to \Iiqt~ over larger times ($10$ ns) are significantly smaller than those obtained from fits over shorter time ($1$ ns) (compare Figs. \ref{fig:f3}b  and \ref{fig:f7}b). The values of $\beta$ are similar for the two simulation times.

\begin{figure}[h]
\includegraphics[scale=0.3,angle=0]{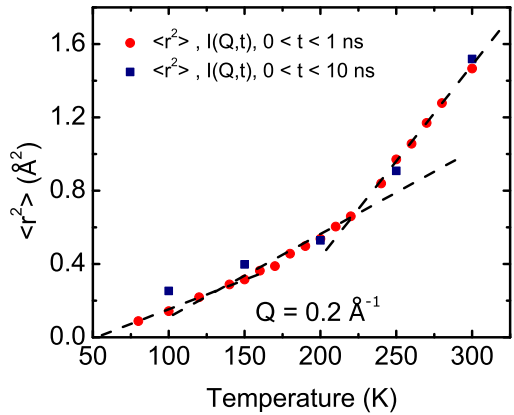}
\vspace{0.5cm}
\caption{The intrinsic MSD \rs~ versus temperature obtained from fits to the \Iiqt~  at $Q = 0.2$ \AA $^{-1}$ obtained from (1) $100$ ns (solid circles) and (2) $1$ $\mu$s MD simulations (solid squares). The \rs~ is largely independent of the simulation time fitted.}
\label{fig:f8}
\end{figure} 

\begin{figure}[h]
\includegraphics[scale=0.3,angle=0]{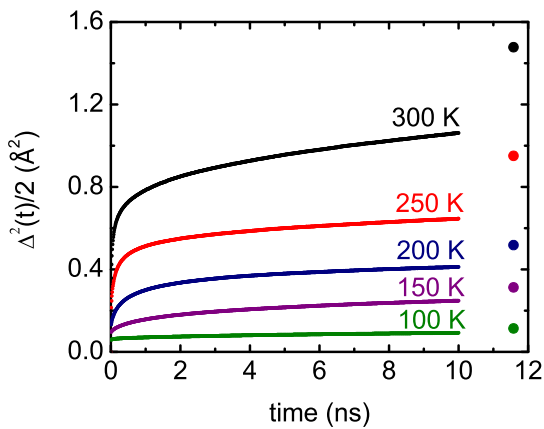}
\vspace{0.5cm}
\caption{The MSD \rssim~ defined in Eq. (\ref{e2}) of non-exchangeable hydrogen versus time at five different temperatures, $100$ K to $300$ K. The dots are the corresponding intrinsic MSD \rs~ at each temperature.}
\label{fig:f9}
\end{figure} 
\begin{figure}[h]
\includegraphics[scale=0.3,angle=0]{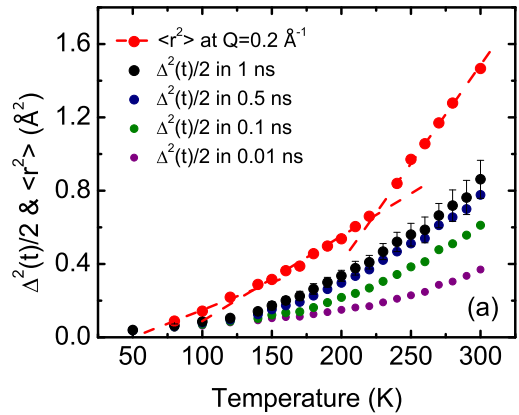}\\
\includegraphics[scale=0.3,angle=0]{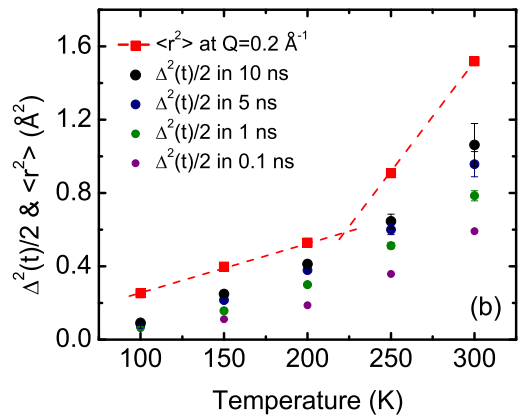}
\vspace{0.5cm}
\caption{Comparison of the intrinsic MSD \rs~ at $Q =0.2$ \AA $^{-1}$ and the MSD \rssim~ out to times (a) $0.01$, $0.1$, $0.5$ and $1$ ns obtained from the $100$ ns MD simulation and (b) out to times $0.1$, $1$, $5$ and $10$ ns, obtained from the $1$ $\mu$s MD simulation.}
\label{fig:f11}
\end{figure}   
\begin{figure}[h]
\includegraphics[scale=0.3,angle=0]{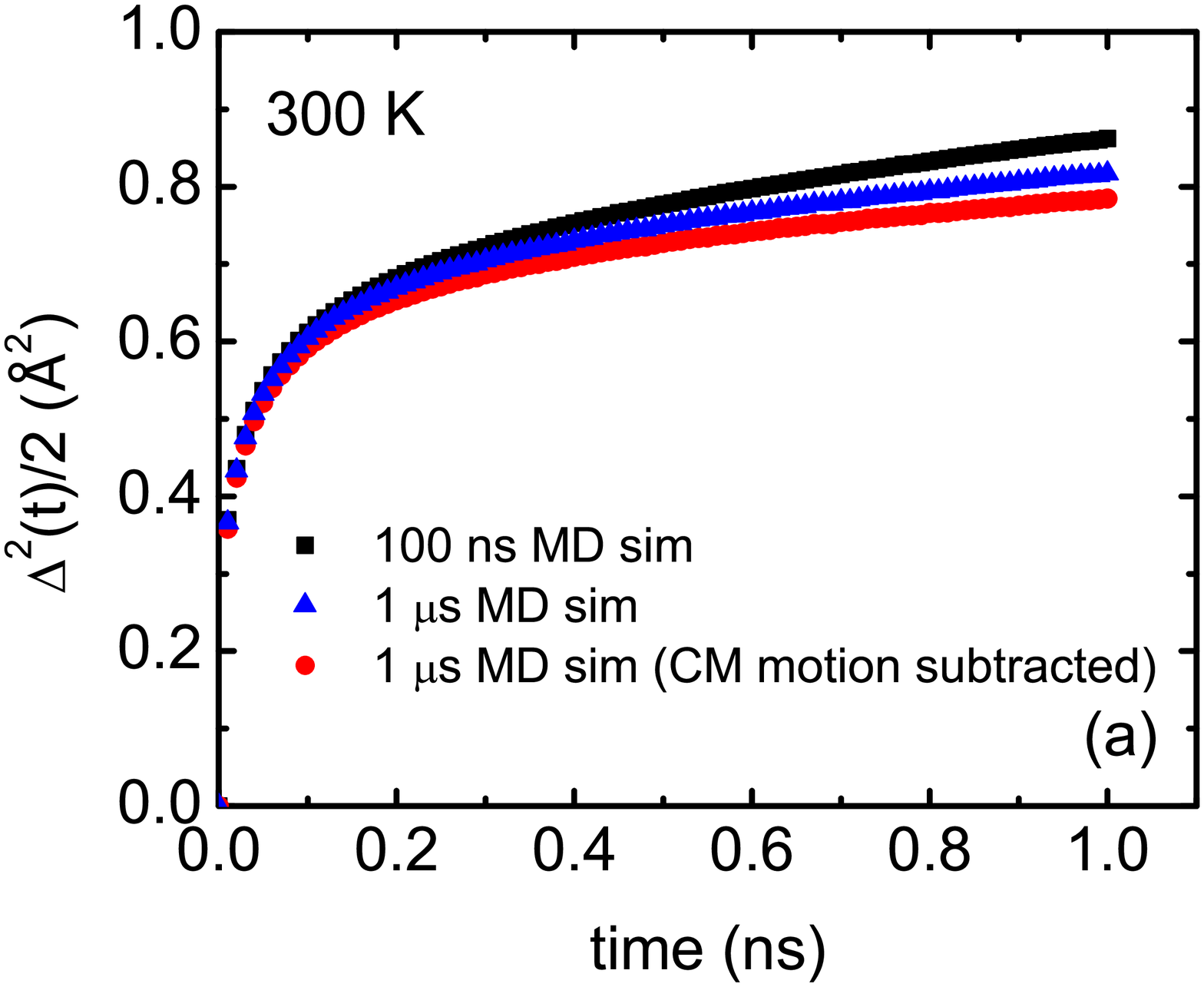}\\
\includegraphics[scale=0.3,angle=0]{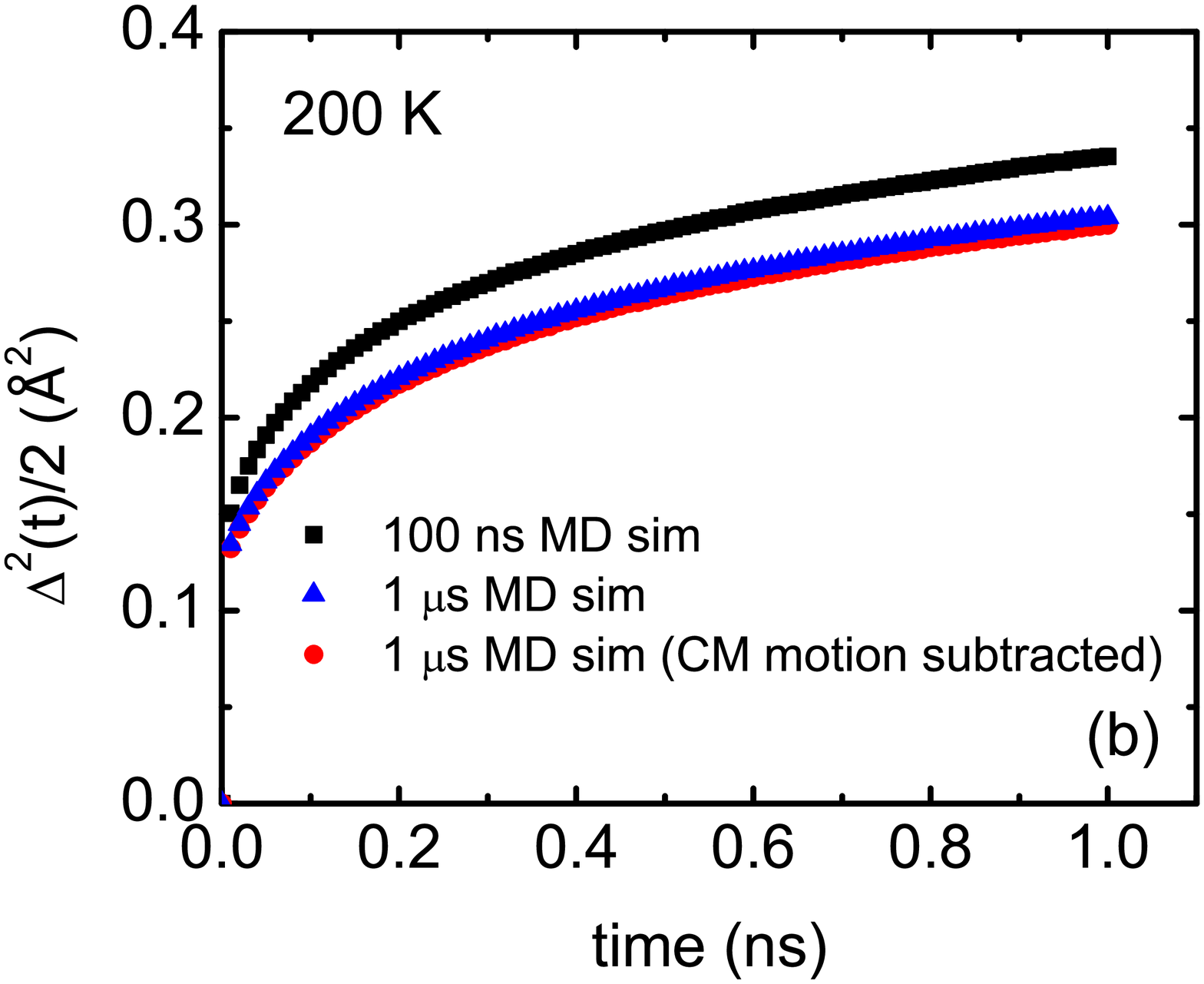}
\vspace{0.5cm}
\caption{The MSD \rssim~ of non-exchangeable hydrogen versus time up to $1$ ns as calculated from MD simulations of $100$ ns (black squares), $1$ $\mu$s (red circles) with and (blue triangles) without CM motion subtracted at (a) $300$ K and (b) $200$ K.}
\label{fig:f10}
\end{figure} 

Fig. \ref{fig:f8} compares the values of $\langle r^2\rangle$ obtained from fits to $I_{inc}(Q,t)$ for times $t$ out to $1$ ns and $10$ ns. The agreement of the two is good and excellent at higher temperatures. This suggests that $\langle r^2\rangle$ is, indeed, a long time ($t\rightarrow\infty$) intrinsic value of $\langle r^2\rangle$ that is independent of the time interval of the data from which it is obtained. It also indicates that no new motional process enters the simulations between $100$ ns and $1$ $\mu$s.

\subsection{Simulated MSD \dsq}

In this section we present values of \dsq~ defined by Eq. (\ref{e2}) and calculated directly using the $r_i(t)$ generated in simulations. The sum in Eq. (\ref{e2}) is taken over the non-exchangeable H nuclei only in the lysozyme. That is, all other nuclei in the protein, the H in the hydration water and the H in the protein that can exchange positions with H in the hydration water (the exchangeable H) are excluded from the sum.
The time dependence of $\Delta ^2(t)$ and estimated values of $\Delta ^2(t = \infty)$ are compared the intrinsic MSD \rs~ defined in Eq. (\ref{e1a}). The \rs~ and $\Delta ^2(\infty)/2$  will be the same only if all H are in identical environments in the protein. Specifically, the two will differ when there is dynamical heterogeniety. 

Fig. \ref{fig:f9} shows $\Delta ^2(t)/2$ obtained from the 1 $\mu$s simulation calculated out to $t = 10$ ns. At $T = 100$ K the \dsq~ appears to have converged after 10 ns and $\Delta ^2(t = 10~\mathrm{ns})$ and \rs~ are quite similar. The \rs~ shown are those from the 100 ns simulations. The \rs~ from the 1 $\mu$s simulations are slightly larger at 100 K and 150 K (see Fig. \ref{fig:f8}). In contrast, at higher temperatures,  $\Delta ^2(t)$ has clearly not reached its terminal ($t = \infty$) value after $10$~ns. For example, at $T = 300$ K, the intrinsic $\langle r^2\rangle$ is approximately $30 - 40$ \% larger than $\Delta ^2(t = 10 \mathrm{ns})/2$. This comparison between $\Delta ^2(t)/2$ and \rs~ is consistent with $\langle r^2\rangle$ representing the intrinsic ($t = \infty$) MSD. 

Fig. \ref{fig:f11} further compares the intrinsic MSD \rs~ at $Q = 0.2$ \AA$^{-1}$ and the $\Delta ^2(t)/2$ at different times $t$ obtained from the 100 ns (Fig. \ref{fig:f11}a) and 1 $\mu$s (Fig. \ref{fig:f11}b) simulations. From Fig. \ref{fig:f11} we see that $\Delta ^2(t)$ has not converged to a long time value after $t = 10$ ns except possibly at 100 K. This is especially true at higher temperature. For example, at $300$ K the increase in  $\Delta ^2(t)/2$ between $1$ ns and $10$ ns is approximately the same as between $0.1$ ns and $1$ ns suggesting that convergence is very slow. The  $\Delta ^2(t)/2$ lie well below the intrinsic $\langle r^2\rangle$, especially at high temperature.

Fig. \ref{fig:f10} shows the $\Delta ^2(t)$ at 200 K and 300 K over the time range $0 < t < 1$ ns as calculated from the $100$ ns simulation and the $1$ $\mu$s simulation. The $\Delta ^2(t)$ obtained from data taken out to $1$ $\mu$s is somewhat smaller than that from the $100$ ns simulation. This suggests that there may be a structural change in the time scale between $100$ ns and $1$ $\mu$s. However, these small differences do not appear to affect \Iiqt~ nor the fitted  intrinsic \rs~ significantly.

Values of $\Delta ^2(t)$ obtained from the $1$ $\mu$s simulation with and without the CM motion subtracted are also shown. The contribution of the CM motion to $\Delta ^2(t)$ is small. Values of $r_i(t)$ corrected for CM motion were used to calculate \Iiqt.

\section{Discussion}

The aim of the present paper is to propose a method to obtain the intrinsic, long time MSD in proteins from finite time simulations. The intrinsic MSD \rs~ is defined as the \rs~ that appears in the infinite time limit of the incoherent intermediate scattering function (ISF) given by Eq.~(\ref{e1a}), often referred to as the Debye-Waller factor. 
The method consists of calculating the ISF \Iiqt~ from a simulation and fitting a model \iqt~ which 
contains \rs~ to the calculated \Iiqt. 
The resulting intrinsic \rs~ exhibits two interesting features: (1) the \rs~ is independent of the simulation time used to calculate \Iiqt, at least up to 1 $\mu$s and (2) the \rs~ shows a clear break in slope of \rs~ vs 
$T$ at the dynamical transition (DT) and a second break at a lower temperature, $T~\simeq$  140 K. The intrinsic \rs~ shows the same breaks in slope that are found in time limited MSD and observed in experiments. This suggests that a DT is an intrinsic property of proteins, not simply an artifact of finite instrument resolution and limited time windows.

\begin{figure}[h]
\includegraphics[scale=0.3,angle=0]{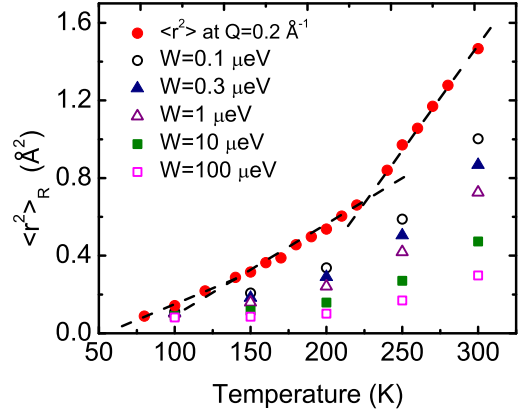}
\vspace{0.5cm}
\caption{The present intrinsic MSD \rs~ at $Q = 0.2$ \AA $^{-1}$ and resolution broadened MSD $\langle r^2\rangle_{R}$ calculated from the same model for the energy resolution widths, $W$, $0.1$, $0.3$, $1$, $10$ and $100$ $\mu$eV.}
\label{fig:f12}
\end{figure} 

\begin{figure}[h]
\includegraphics[scale=0.3,angle=0]{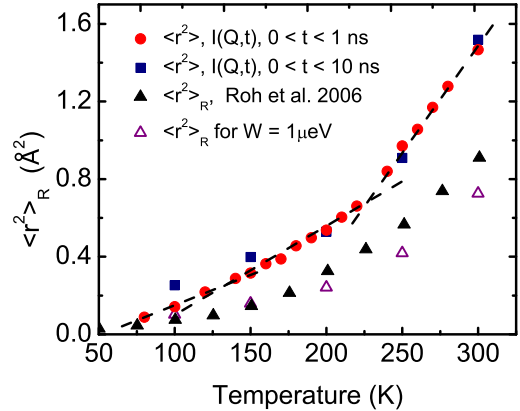}
\vspace{0.5cm}
\caption{The present intrinsic MSD \rs~ at $Q = 0.2$ \AA $^{-1}$ (obtained from fits to (1) $100$ ns (solid squares) and (2) $1$ $\mu$s MD simulations (solid circles)) and the present resolution broadened MSD $\langle r^2\rangle_{R}$ ($W = 1$ $\mu$eV)(open triangles) for lysozyme at h = 0.40 compared with simulated \rsr~ at $W$ = 1 $\mu$eV by Roh \textit{et al.} 2006 \cite{Roh:06} for lysozyme at h = 0.43 (solid triangles).}
\label{fig:f13}
\end{figure} 

\begin{figure}
\begin{center}
\includegraphics[scale=0.3,angle=0]{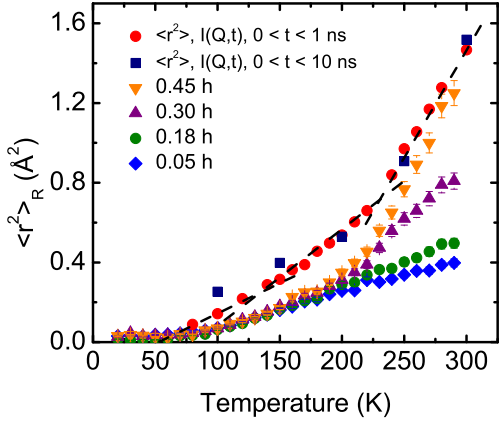}
\end{center}
\vspace{0.5cm}
\caption{The present intrinsic MSD \rs~ at $Q = 0.2$ \AA $^{-1}$ for lysozyme (h = 0.4), as in Fig.~\ref{fig:f13}, compared with the experimental resolution broadened MSD \rsr~ for $W$ = 1 $\mu$eV for lysozyme at different hydration levels (h) observed by Roh \textit{et al.} 2006 \cite{Roh:06}.}
\label{fig:f14}
\end{figure}


\subsection{Comparison with existing MSD for lysozyme}

To place the present intrinsic \rs~ in context with existing MSD, we firstly compare \rs~ with resolution broadened MSD \rsr~ both calculated from the present simulations of lysozyme. A resolution broadened \rsr~ is an MSD that has developed over a finite time only. This time, \tr, is determined by the resolution width as \tr~= $(8ln2)^{1/2} \hbar/W$ for a Gaussian resolution function. The smallest FWHM readily available today is $W$ = 1 $\mu$eV ($\tau_R $ = 1.5 ns). By using the same simulation for both \rs~ and \rsr, we can isolate the impact of a finite resolution width. 

Simulated and observed MSD are usually compared by comparing resolution broadened \rsr, MSD that have developed over the same time period.\cite{Tarek:00a,Roh:06,Wood:08} Excellent agreement between simulated and observed MSD has been obtained in this way. Specifically, the resolution broadened MSD, \rsr~ is obtained from Eq.~ (\ref{e1}) in which \srqw~ is the observed, resolution broadened DSF,
\begin{equation} \label{e13}
S_R(Q,\omega) = \frac{1}{2\pi} \int \mathrm{d}t \exp(i\omega t) I(Q,t)R(t).
\end{equation}
and $R(t)$  is Fourier transform of the instrumental resolution function in time. $R(t)$ is typically a Gaussian, $R(t) = \exp(- \frac{t^2}{2\tau_R ^2})$. The resolution function cuts off \iqt~ after a time \tr~= $(8ln2)^{1/2} \hbar/W$. In experiment, the observed \rsr~ is obtained by inserting the observed \srqwe~ in Eq.~ (\ref{e1}). In simulations the calculated ISF is inserted in Eq.~(\ref{e13}) to obtain \srqwe~ and then \rsr~ is again obtained using Eq.~(\ref{e1}). In this way an \rsr~ that has evolved over to a time \tr~ is compared. We follow exactly this procedure to calculate \rsr~ from the present simulations by substituting our model \iqt~ into Eq.~(\ref{e13}).

The intrinsic \rs~ and the resolution broadened \rsr~ of lysozyme are compared in Fig.~\ref{fig:f12}. The \rs~ is found to be approximately twice the \rsr~ for a resolution width $W$ = 1 $\mu$eV (\tr~= 1.5 ns).  The ratio \rs/\rsr~ is approximately independent of temperature for $T >$ 150 K. The \rs~ remains well above \rsr~ for $W$ = 0.1 $\mu$eV (\tr~= 15 ns), a resolution approximately ten times higher than that available today. Physically, we expect resolution broadening to reduce \rsr~ below \rs~ if \tr~ is less than or comparable to the longest relaxation time $\tau = \lambda^{-1}$ of the protein. From Figs.~\ref{fig:f3}  and \ref{fig:f7} we see that, at low $Q$ and temperatures above 170 K,  $\tau = \lambda^{-1} \ge $ 1 ns. The $\tau$ is somewhat longer at low temperature.  On this basis we expect resolution broadening to be important at $W$ = 1 $\mu$eV. The degree of broadening depends sensitively on the functional form of $C(t)$ in the model \iqt, as discussed below.

Next we compare the present simulated \rsr~ with previous simulated values of \rsr~ for lysozyme. Roh  \textit{et al.}\cite{Roh:06} have calculated \rsr~ at low $Q$ and $W$ = 1 $\mu$eV from their simulations of lysozyme hydrated to h = 0.43. The Roh \textit{et al.} \rsr~ and the present \rsr~ at low $Q$ ($Q$ = 0.2 \A) and $W$ = 1 $\mu$eV for h = 0.40 are compared in Fig.~\ref{fig:f13}. The agreement is excellent since the present \rsr~ is somewhat lower as expected since the present h is lower and the MSD is very sensitive to h.
 
Thirdly, we compare the present intrinsic \rs~ with \rsr~ observed experimentally in lysozyme using an energy resolution width $W$ = 1 $\mu$eV. Observed \rsr~ of lysozyme at four hydration levels and the present \rs~ are shown in Fig.~\ref{fig:f14}. The observed \rsr~ are very sensitive to the hydration level for $T >$ \td. In  Fig.~\ref{fig:f14} the present intrinsic \rs~ for h = 0.4 lies above but close to the observed \rsr~ for h = 0.45 and significantly higher than the observed \rsr~ for lower hydrations, as expected.  Comparing the calculated \rsr~ in Fig.~\ref{fig:f13} with the observed \rsr~ in Fig.~\ref{fig:f14}, we see that the calculated \rsr~ lie somewhat below but close to the experimental values for similar levels of hydration. Broadly the agreement between the simulated and observed \rsr~ is very good, both in terms of the absolute value and in the temperature dependence. 

\subsection{Sensitivity of \rs~ to the model $C(t)$ }

\begin{figure}[h]
\includegraphics[scale=0.3,angle=0]{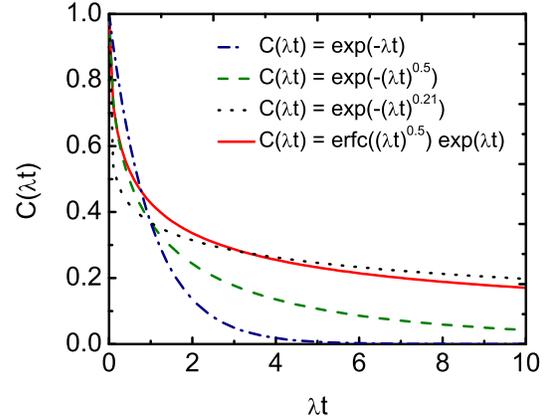}
\vspace{0.5cm}
\caption{Comparison of relaxation functions: a simple exponential (blue dashed dotted line), a stretched exponential with $\beta = 0.5$ (green dashed line), a stretched exponential with $\beta = 0.21$ (black dotted line) and the Mittag-Leffler function (red solid line).}
\label{fig:f15}
\end{figure}

The difference between the intrinsic \rs~ and the resolution broadened \rsr, such as shown in Fig.~\ref{fig:f13}, is sensitive to the functional form of $C(t)$ used to describe the dynamic correlations in the protein in the model \iqt. Four forms of C(t) are compared in Fig.~\ref{fig:f15}. The stretched exponential function for $C(t)$ given by Eq.~(\ref{e10}) that we have used here provides a reasonable fit of the model \iqt~ to the calculated \Iiqt~ provided the parameter $\beta$ is small, i.e. $\beta$ = 0.23. In Fig.~\ref{fig:f15} we see that the stretched exponential for a small $\beta$ and the Mittag-Leffler function\cite{Calandrini:08} have long range tails reaching out to times a factor of ten beyond  t = $\lambda^{-1}$. This means that correlations persist for times well beyond $\tau = \lambda^{-1}$. For this reason when a stretched exponential with a small $\beta$ is used we expect \rs~ to lie above \rsr~ even when $\lambda^{-1} \simeq$ \tr, as found here in Fig.~\ref{fig:f13}. In contrast if $C(t)$ is represented by a simple exponential, or a stretched exponential with a large value of $\beta$, the correlations die out rapidly on a time scale $\tau = \lambda^{-1}$.  It was not possible to obtain a good fit to \Iiqt~ using a simple exponential or a large $\beta$. Thus correlations that persist to long times  t = 10 $\lambda^{-1}$ appears to be a feature of lysozyme. 

In an earlier study\cite{Vural:12}, we proposed a method to obtain the intrinsic MSD in proteins from fits to experiment, to observed resolution broadened \srqwe. The model \iqt~ employed was the same as that used here in Eq.~(\ref{e11}). The model \iqt~ was Fourier transformed (see  Eq.~(\ref{e13})) to obtain \srqwe. 
However, $C(t)$ was represented by a simple exponential, chosen because the experimental data were not very discriminating and fits using a simple exponential and a stretched exponential could not be distinguished. The ratio \rs/\rsr~ obtained from fits to data at $W$ =1 $\mu$eV was approximately 1.0 - 1.2 rather than a factor of two as found here. MD simulation-derived \Iiqt~ are more discriminating. We believe that the present $C(t)$ and ratio  \rs/\rsr~ are more accurate because the $C(t)$ obtained from simulations is more accurate. Hence, the method proposed in Ref. (\onlinecite{Vural:12}) to obtain \rs~ from experiment needs to be upgraded by replacing the exponential $C(t)$ by a stretched exponential with $\beta$ set at approximately 0.23.

The model could conceivably be further refined by using more sophisticated expressions for $C(t)$ that combine vibrational motion at short times and diffusion at longer times. Also the expression for the time dependent part of \iqt~ in  Eq.~(\ref{e9}) may be too simple. 

\subsection{Dynamical Transition and Impact of Instrument Resolution}

\begin{figure}[h]
\includegraphics[scale=0.3,angle=0]{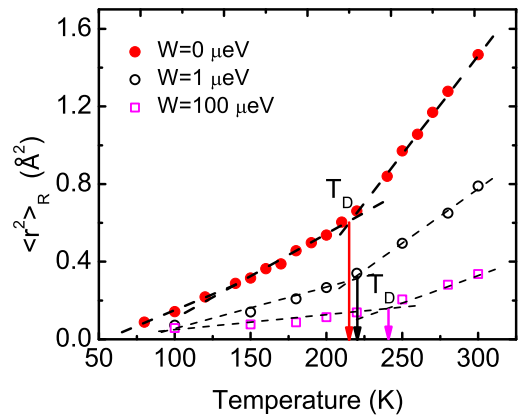}
\vspace{0.5cm}
\caption{The intrinsic MSD ($W$ = 0) and resolution broadened MSD \rsr~ for $W$ = 1 $\mu$eV and $W$ = 100 $\mu$eV obtained from Eqs. (\ref{e1}) and (\ref{e13}) using the present model \iqt~ (reproduced from Fig.~\ref{fig:f12}) with the dynamical transition temperature \td~ identified.}
\label{fig:f16}
\end{figure}
\begin{figure}[h]
\includegraphics[scale=0.3,angle=0]{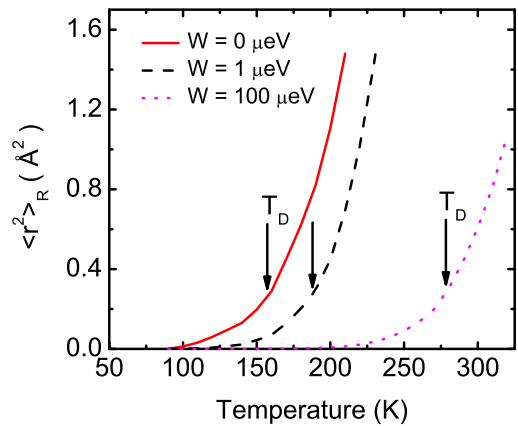}
\vspace{0.5cm}
\caption{The intrinsic MSD ($W$ = 0) and resolution broadened MSD \rsr~ for $W$ = 1 $\mu$eV and $W$ = 100 $\mu$eV with \td~ identified. The MSD are obtained from fits to experiment in Ref. \onlinecite{Vural:12} using a model \iqt~ that has a simple exponential decay function $C(t) = \exp[-\lambda t]$.}
\label{fig:f18}
\end{figure} 

Despite decades of experimental and theoretical studies, the physical origin of the dynamical transition (DT) remains debated. It has been ascribed\cite{Khodadadi:10} to sudden change of “effective elasticity” in proteins\cite{Zaccai:00}, to the onset of motions of specific side groups, e.g., methyl group rotations,\cite{Lee:01} to a glass transition or a phase transition in the hydration water\cite{Doster:11,Chen:06}, and interpreted as an apparent effect arising because the MSD is observed with a finite instrument resolution width.\cite{Daniel:99,Becker:04,Khodadadi:08,Fenimore:04}   
In this section we discuss the impact of observing the DT using an instrument having a finite energy resolution width, W, within the present model. Firstly, the intrinsic MSD \rs~ that we have found in this paper for lysozyme and is shown in Figs.~\ref{fig:f5} and \ref{fig:f8} displays a clear DT at a transition temperature, \td, of 220 K. This result suggests that, within the rigor of simulations, the DT is an intrinsic property of a protein. The DT is not simply an artifact of observing the MSD with an instrument having a finite $W$ and a limited time window. However, the change of slope of the MSD at \td~ can be modified and \td~ shifted to higher temperature when the DT is observed with a finite $W$.\cite{Nakagawa:10,Jasnin:10,Wood:08,Daniel:03} as emphasized recently.\cite{Schiro:12} 

When $W$ is finite, the MSD \rsr~ defined above in Eqs. (\ref{e1}) and (\ref{e13}), rather than \rs~ is observed. In Eq. (\ref{e13}), motions in the protein can contribute to the \rsr~ for a limited time $\tau_R \simeq \hbar/W$ only.  Since motions over a limited time window $\tau_R$ are included,  \rsr~ is always smaller than \rs~ at a given temperature. As a result the \td~ in \rsr~ is shifted to a higher temperature. Using the present model, this shift is illustrated in Fig. \ref{fig:f16}  where the \td~ is explicitly identified. For example, when observed with an instrument for which W = 100 $\mu$eV (\tr~ = 15 ps for a Gaussian resolution function), the apparent \td~ is shifted to 240 K.
 
The degree of impact of $W$ on \rsr~ depends on the rate at which correlations decay in the protein. In the present model, the decay rate depends on the magnitude of the parameter $\lambda$ and on the functional form of $C(t)$. The present stretched exponential $C(t)$ has a long time tail (see Fig.~\ref{fig:f15}). This means that the reduction of \rsr~ below \rs~ begins at small values of W (long \tr). But the rate of change of \rsr~ with $W$ is gradual. In an earlier model\cite{Vural:12}, $C(t)$ was  described by a simple exponential which falls rapidly with $t$ (see Fig.~\ref{fig:f15}). For this $C(t)$,  the reduction of \rsr~  below \rs~ begins at a larger value of $W$ and thereafter the reduction increases rapidly with increasing $W$. We reproduce the \rsr~ obtained for an exponential $C(t)$ in Fig.~\ref{fig:f18}. The \td~ increases rapidly with increasing $W$ which illustrates the dependence of \td~ on $W$ vividly. Using this simple model, the increase of the apparent \td~ with increasing $W$ can also be readily understood. In the model, \rsr/\rs~ = $[1+\frac{W}{I_{\infty} \lambda}]^{-1}~ \simeq~[1-\frac{W}{I_{\infty} \lambda} + ...]$.  The parameter $\lambda$ increases with increasing $T$. Thus for a given $W$, the ratio $W/(I_{\infty}\lambda)$ decreases with increasing temperature and \rsr/\rs~ is larger at higher temperature. Thus \rsr~ is decreased least by finite resolution at the highest temperatures. 

As illustrated by these models, a DT is readily observed on an instrument having a finite resolution width. The chief impact of a finite $W$ is to shift the apparent \td~ to a higher temperature.  

\subsection{MSD calculated from Simulations} 

\begin{figure}[h]
\begin{center}
\includegraphics[scale=0.3,angle=0]{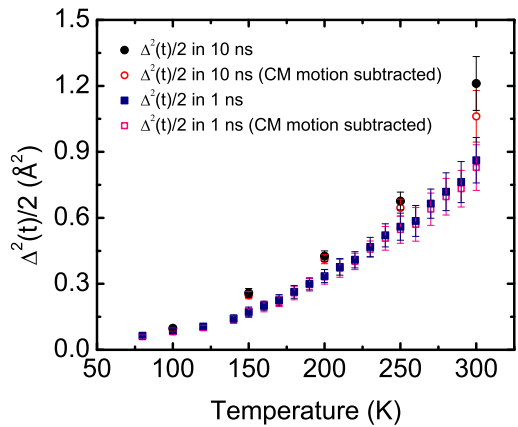}
\end{center}
\vspace{0.5cm}
\caption{The MSD \rssim, with and without the CM motion subtracted after $t = 1$ ns ($100$ ns MD simulation) and $t = 10$ ns ($1$ $\mu$s MD simulation).}
\label{fig:f17}
\end{figure}

We have also evaluated the MSD \dsq~ for H in lysozyme defined in Eq. (\ref{e2}) from the present simulations. The \dsq/2 will be the same as the intrinsic \rs~ only if (1) the \dsq~ has reached its long time, converged value so that the correlations are zero as discussed in Eq.~(\ref{e3}), and (2) if all the H in the protein are in identical environments so that \Iiqt~ reduces to the model \iqt. Also, in the present work, only the non-exchangeable H nuclei were included in \dsq~ while the intrinsic \rs~ is obtained from a fit to an \Iiqt~ which includes all nuclei. We expect the last difference to be minimal and it would be straightforward to include only the H nuclei in \Iiqt~ if desired. With these caveats, we have compared the \dsq/2~ with \rs~ in  Figs.~\ref{fig:f9} and \ref{fig:f11}. At low temperature (e.g. 100 K) where diffusion is expected to be less important, we find that \dsq/2~ appears to have converged after 10 ns and approaches \rs~ reasonably well. However, at higher temperature (e.g. 250 K), \dsq/2~ has not converged to a constant after 10 ns and lies well below \rs. From these comparisons it would be interesting to evaluate \dsq~ out to longer times to determine whether it converges and to reveal the dynamics contributing. For example, at 300 K, nearly translational diffusion may be possible for some H in the protein that are near the surface or near hydration water. It would be interesting to exclude these H from \dsq. In this regard it is also important to exclude the CM motion which becomes important at higher temperature and longer times as shown in Fig.~\ref{fig:f17}.

\section{Conclusion}

We have proposed a procedure to obtain the intrinsic, long time MSD in proteins from finite time simulations. The intrinsic MSD represents the equilibrium MSD as would be predicted by statistical mechanics and the energy landscape, assuming the protein does not go through major structural changes. The specific MSD investigated is the one determined in neutron scattering measurements. The intrinsic MSD is calculated from simulations of 100 ns and 1 $\mu$s and found to be independent of simulation time. The intrinsic, long time MSD in lysozyme is found to be approximately twice the MSD that develops after a time of 1.5 ns, as would be observed using neutron instruments with an energy resolution width of $W$ = 1 $\mu$eV. The intrinsic MSD shows the same breaks in slope with temperature as does the finite time MSD. The ratio of the intrinsic to finite time MSD is sensitive to the model functions (e.g.  stretched exponentials) used to describe the motions in the protein as well as to the decay times of the motions themselves.   

\section{Acknowledgements}

It is a pleasure to acknowledge valuable discussions with Mark Johnson, Giuseppi Zaccai and Dominique Bicout. This work was supported by the DOE, Office of Basic Energy Sciences, under contract No ER46680 (DV and HRG) and by NSF grant number MCB-0842871 (LH and JCS). 


\end{document}